\documentclass{aa}  
\usepackage{graphicx}
\usepackage{txfonts}
\usepackage[]{hyperref}
\usepackage{xspace} 
\newcommand\rebound{{\texttt{REBOUND}}\xspace}
\providecommand{\e}[1]{\ensuremath{\times 10^{#1}}}
\usepackage{siunitx}
\usepackage{amstext}

\begin{document}

   \title{Investigating gravitational collapse of a pebble cloud to form transneptunian binaries}


   \author{J. E. Robinson \inst{1}
          \and
          W. C. Fraser \inst{1, 2} 
          \and
          A. Fitzsimmons \inst{1}
          \and
          P. Lacerda \inst{1}
      }

   \institute{Astrophysics Research Centre, Queen's University Belfast,
              University Road, BT7 1NN, Belfast, Northern Ireland\\
              \email{j.e.robinson@qub.ac.uk}
         \and
             Herzberg Astronomy and Astrophysics Research Centre, National Research Council of Canada, 5071 West Saanich Road, Victoria BC V9E 2E7, Canada
             }

   \date{Received January ??, 2020; accepted ?? ?? ??}

 
  \abstract
{A large fraction of transneptunian objects are found in binary pairs, $\sim30\%$ in the cold classical population between $a_\mathrm{hel}\sim 39$ and $\sim48\, \mathrm{AU}$. Observationally, these binaries  generally have components of similar size and colour. Previous work has shown that gravitational collapse of a pebble cloud is an efficient mechanism for producing such systems. Since the bi-lobate nature of 2014 MU$_{69}$ (Arrokoth) was discovered,   interest in gravitational collapse as a pathway for forming contact binaries has also grown.}
{We investigate the formation of binary systems through gravitational collapse by considering a wider range of binary masses than previous studies.
        We analysed in detail the properties of the bound systems that are formed and compared them to observations.}
   {We performed $N$-body simulations of gravitational collapse of a pebble cloud using the \rebound package, with an integrator designed for rotating reference frames and robust collision detection. We conducted a deep search for gravitationally bound particles at the end of the gravitational collapse phase and tested their stability. For all systems produced, not just for the most massive binaries, we investigated the population characteristics of their mass and orbital parameters.}
   {We found that gravitational collapse is an efficient producer of bound planetesimal systems. On average, $\text{about } 1.5$ bound systems were produced per cloud in the mass range studied here. In addition to large equal-sized binaries, we found that gravitational collapse produces massive bodies with small satellites and low-mass binaries with a high mass ratio. Our results disfavour the collapse of high-mass clouds, in line with reported upper mass limits of clouds formed by the streaming instability. Gravitational collapse can create binary systems analogous to Arrokoth, and collisions in a collapsing cloud should be gentle enough to preserve a bi-lobed structure.
   }
   {}


\titlerunning{Formation of transneptunian binaries}

   \maketitle

\section{Introduction}

The transneptunian objects, also known as Kuiper belt objects, are the mainly icy Solar System bodies beyond the orbit of Neptune. 
They are generally classified by their heliocentric orbits according to the taxonomy laid out by \cite{Gladman2008}.
The main dynamical classes are cold classicals, hot classicals, resonant objects, scattered disc, and fossilized scattered disc.
A classical object is a transneptunian object that is not under the influence of or has been scattered by Neptune. 
Most classicals are found on orbits between the 3:2 and 2:1 Neptune resonances.
In this case, `cold' refers to objects on heliocentric orbits that are less dynamically excited (they have lower eccentricity and inclination) than the `hot' objects.
The transneptunian objects contain some of the most primitive solid material in the Solar System, and their physical properties and dynamical structure provide a valuable insight into Solar System history. 
A particularly interesting property of the transneptunian populations is the fraction of bodies in mutual orbit binary pairs, which varies across the different dynamical classes.
For example, unlike the excited objects of which only $\sim10\%$ are in binary pairs \citep{Stephens2006}, the predominantly red surface colour and dynamically quiescent cold classical objects exhibit a high binary fraction of $\sim30\%$ \citep{Fraser2017,Grundy2011,Noll2008}.

These transneptunian binaries (TNBs) generally have similar-size components \citep{Noll2008}, both of which have similar surface colours \citep{Benecchi2009,Marsset2020}.
Furthermore, some TNBs have wide orbital separations relative to the mutual Hill radius of the binary ($a_\mathrm{bin}/R_\mathrm{Hill}>0.05$, where $a_\mathrm{bin}$ is the binary orbit semimajor axis and $R_\mathrm{Hill}$ is the Hill radius of the binary system).
These properties are rather unique, especially when compared to asteroidal binary systems, which generally have smaller secondaries on tighter orbits \citep{Walsh2015}. 
They also display some interesting distributions in the orientation of their binary orbits.
The most recent set of full orbital solutions \citep{Grundy2019} shows that the tight binaries with $a_\mathrm{bin}/R_\mathrm{Hill}<0.05$ have a majority of prograde orbits (22 out of 26), whereas the wide binaries have a \textit{\textup{somewhat}} higher fraction of retrograde systems (6 out of 9 are prograde).
However, we note that assuming Poisson statistics, the number of data points for the wide systems is too low to definitively state that their retrograde fraction is different from that of the tight binaries.
It is clear, however, that the wide systems tend not to be highly inclined relative to their heliocentric orbits compared to the tight systems, which exhibit a full range of inclinations \citep{Parker2011,Grundy2011,Grundy2019}.
These unique properties present the opportunity of providing excellent constraints for theories of planetesimal and binary formation, and the subsequent evolution of the small-body population from the early Solar System to the present day.
\\

In this work we focus on the formation of binary systems that are representative of the TNBs.
The current literature knows two main classes of formation mechanisms: hierarchical coagulation and gravitational instability.
The first has many sub-mechanisms, but they involve in general two or more planetesimals interacting with some loss of energy, which results in two bodies remaining in a bound pair.
Some proposed mechanisms for removing excess energy include the ejection of a third body (L$^3$), dynamical friction with background small bodies (L$^2$s) \citep{Goldreich2002}, or collisions \citep{Weidenschilling2002,Funato2004}. 
These methods can explain some properties of the TNBs, but may either be too inefficient, require unrealistic protoplanetary disc conditions, or fail to reproduce all the observed binary properties, for instance\ same colour components or the distribution of the binary orbit inclination.

The second mechanism, gravitational instability, was originally proposed as a method of forming planetesimals that overcomes the various barriers that oppose the growth of protoplanetary disc dust grains into larger bodies, for example the bouncing, fragmentation, and radial-drift barriers (see \cite{Weidenschilling1977,Brauer2008,Guttler2010,Birnstiel2016}).
In this mechanism, solid particles in the gaseous protoplanetary disc are concentrated by the streaming instability \citep{Youdin2005} until the mutual self-gravity of a cluster is high enough for the cluster to collapse.
During this collapse the particles collide and merge, allowing the rapid growth of $\sim 100\, \si{km}$ planetesimals whilst overcoming the barriers above \citep{Johansen2007,Johansen2015,Simon2016}.
This mechanism was first proposed as a possible route to TNB formation by \cite{Nesvorny2008} and \cite{Nesvorny2010}.
The particle clumps formed by streaming instability generally have some rotation as a result of vorticity in the gaseous protoplanetary disc. 
This means that it is difficult to form a single object without removing angular momentum from the particle cloud.
Instead it is more probable for the cloud to collapse and transfer its angular momentum into a binary planetesimal system. 
For a given binary orbit the angular momentum of the binary is maximised when the components have equal mass.
The streaming instability has been shown to be robust across a range of protoplanetary disc conditions, generally requiring a solid-to-gas density 2-5 times that of the solar ratio, see \cite{Johansen2009,Yang2017} and also \cite{Carrera2017,Nesvorny2019}.

The gravitational collapse of pebble clouds formed by streaming instability is extremely efficient at forming binary systems with the unusual properties of the TNBs: similar-size components, wide binary orbits, and similar colours (as the bodies formed simultaneously from the same cloud of material).
Moreover, the inclination of a binary formed by gravitational collapse of a pebble cloud is dependent on the inclination of the cloud from which it forms.
The clouds formed by streaming instability have a distribution of prograde and retrograde rotations, which means that binary formation by gravitational collapse can reproduce the observed binary inclination distribution \citep{Nesvorny2019}.
\\

Since the original work by \cite{Nesvorny2010}, who proposed gravitational collapse as a binary formation mechanism of TNBs, there has been a significant amount of research that supports this theory.
For example \cite{Fraser2017} investigated the population of blue surface colour objects amongst cold classicals, nearly all of which exist as dynamically fragile, widely separated binary pairs.
Different surface colours imply a different composition and therefore a different formation region.
Despite being widely separated, \cite{Fraser2017} showed that these blue binaries could survive being pushed outwards from their formation region by the migration of Neptune.
The implication of this study is that the planetesimals that formed in the same region as these blue binaries must all have formed as binary systems.
This is strong evidence for the formation of planetesimals and binaries through gravitational collapse, as only gravitational collapse is capable of  producing a $\sim 100\%$ binary fraction in addition to producing binaries with the properties discussed above.
        
By analysing high-resolution hydrodynamic simulations, \cite{Nesvorny2019} were able to study the properties of pebble clumps formed by streaming instability in detail.
These clumps would then undergo gravitational collapse and can efficiently form binary systems. 
The authors showed that the orientations of these clumps matched the observed orientations of the TNBs, in particular, the 4:1 preference for prograde binary orbits as shown by \cite{Grundy2019}.
Only the combination of streaming instability and gravitational collapse can form binaries with this particular inclination distribution, thus making it the favoured mechanism to form TNBs and planetesimals.

The New Horizons mission has further stoked interest in the mechanisms of binary formation.
After visiting the Pluto-Charon system, New Horizons performed a flyby of the cold classical transneptunian object, 2014 MU$_{69}$ \citep{Stern2019}, now known as 486958 Arrokoth.
This event was significant as this was the most distant spacecraft flyby to date and the first flyby of a Kuiper belt object.
Being a cold classical object, Arrokoth is dynamically unexcited, and is believed to be composed of primitive material that has not been significantly altered since its formation.
This makes it the perfect tool for probing accretion processes in the early Solar System.
Imaging from the flyby revealed that this small object is composed of two distinct, similar-size lobes in a contact binary configuration.
The implication is that the bi-lobate shape of Arrokoth must be primordial, and as discussed by \cite{Stern2019} and \cite{McKinnon2020}, one of the favoured explanations is the formation of Arrokoth through gravitational collapse. 
This mechanism provides a path for producing bound planetesimals with nearly equal masses in a low-velocity collisional environment.
\\

In this work we build upon the results of \cite{Nesvorny2010}. 
We study the formation of binary systems through gravitational collapse in an independent manner, recreating the simulations within a different numerical framework using the \rebound package.
We perform our own detailed analysis, conducting a deeper search for bound particles, and take an in-depth look at the properties of the binary systems that are formed.
The structure of the paper is as follows: firstly, we describe the setup for our gravitational-collapse simulations using the \rebound package, the search for bound systems, and the dynamical evolution of any systems formed.
We then present our results and perform a detailed analysis of the properties of the binary systems produced by our simulations.
We compare our results to previous work and also to the latest data on the observed TNBs, and discuss the implications and limitations of our work, in particular, the effects of particle-size inflation and timestep-dependent collision detection.

\section{Methods}

\subsection{Simulation with \rebound}

We set up our collapsing cloud simulations with the open source $N$-body package \rebound \citep{Rein2012}.
We used the symplectic epicyclic integrator within the rotating Hill frame, which is a fixed-timestep, second-order mixed variable symplectic method \citep{Rein2011}. 
An oct-tree algorithm \citep{Barnes1986} was used to calculate gravity and to search nearest neighbours for collisions. 
We used the built-in \texttt{openmp} functionality to parallelise and take advantage of the strong scaling of the code.
Collisions were resolved as simple inelastic mergers.
We used open boundary conditions, where particles were removed if they left the simulation box. 
The simulation box was chosen to be sufficiently large such that only particles ejected from the cloud were removed: box size = 20 cloud radii.
The simulation state was saved at regular intervals.
These timestamps allowed us to investigate the process of cloud collapse up until the maximum simulation time of $t=100\, \si{yr}$.

\subsection{Initial conditions}

For our initial conditions we adopted the values in \citet{Nesvorny2010}. 
In summary, a spherical uniform cloud of $N_\mathrm{p}=10^{5}$ particles was defined by the total cloud mass $M_\mathrm{c}$,
with cloud rotation modelled as the rotation of a solid body with angular frequency $\Omega$.
Different rotation states were investigated, defined by the cloud rotation relative to the orbital angular frequency of a particle in a circular orbit at the edge of the cloud $X=\Omega/\Omega_\mathrm{circ}$, where $\Omega_\mathrm{circ}=(GM_\mathrm{c}/R_\mathrm{c}^3)^{1/2}$ and $R_\mathrm{c}$ is the cloud radius. 
Particle size was controlled by an inflation factor $f^*$, such that the modelled particle radius was $r_f=f^*r^*$, where $r^*$ is the radius of a uniform sphere of mass $m_0=M_\mathrm{c}/N_\mathrm{p}$ and density $\rho$.
Throughout this work we have assumed a material density of $\rho=1000\, \si{kg\,m^{-3}}$ to represent icy transneptunian material.
The inflation factor, $f^*$, was included as a means of boosting the collision rate of the simulations and to ensure collisional damping of random velocities during collapse.
Otherwise, given the limited number of computational particles and the therefore low particle number density, the growth of particles by accretion is significantly slower than the collapse timescale.

\citet{Nesvorny2010} explored 240 initial states: cloud masses spanning orders of magnitude $M_\mathrm{c} =4.19\e{18}$, $6.54\e{19}$ , and $1.77\e{21}\, \si{kg}$ \footnote{These cloud masses are equivalent to a spherical body, uniform density $\rho=1000\, \si{kg\,m^{-3}}$, of radius $R_\mathrm{eq}=100$, $250,$ and $750\, \si{km,}$ respectively.}, 
with rotation states ranging from sub- to super-circular rotation $X=0.5,0.75,1.0,\text{and }1.25$, 
and a range of particle inflation factors $f^*=1,3,10,30,\text{and }100$.
Each combination of states above was repeated using four unique particle-position seeds.
We dropped the $f^*=1$ and $X=1.25$ states, so that our dataset has 144 initial states.
We ignored these cloud initial states as \citet{Nesvorny2010} discussed that these parameters are not conducive to forming binary objects.
Our initial investigations of this region of parameter space showed similar trends.
The self-gravity of clouds with $\Omega>\Omega_\mathrm{circ}$ is not enough to overcome their rotation, therefore those clouds do not efficiently collapse and are dispersed due to excess angular momentum.
Particles in clouds with $f^*=1$ have an extremely small collisional cross section, therefore particle growth by merging collisions is minimal over the 100-year timescale we investigated.
\\

The particle positions were generated by randomly and uniformly distributing $N_\mathrm{p}$ particles in the volume enclosed by a unit sphere. These seed positions were then rescaled according to the cloud mass at the start of each simulation, resulting in particle positions $\vec{r}=\left(x,y,z\right)$ distributed in a spherical volume of radius $R_\mathrm{c}=0.6 R_\mathrm{Hill,c}$, where $R_\mathrm{Hill,c}=(GM_\mathrm{c}/3\Omega_\mathrm{Kep}^2)^{1/3}$ is the Hill radius of the cloud at a heliocentric, Keplerian orbital frequency of $\Omega_\mathrm{Kep}$, and the factor 0.6 ensures that all particles start bound within the cloud.

Initial particle velocities, $\vec{v}=(v_x,v_y,v_z)$, are the sum of the rotational velocity of the cloud and the Keplerian shear velocity due to the rotating Hill reference frame,
\begin{equation}
\vec{v}=\vec{v}_\mathrm{rot}+\vec{v}_\mathrm{shear}
.\end{equation}
The angular velocity of the cloud is given by $\Omega=X\Omega_\mathrm{circ}$, and the cloud rotates about the $\hat{\vec{z}}$ -axis, therefore $\vec{\Omega}=X\Omega_\mathrm{circ}\hat{\vec{z}}$.
The rotational velocity component is then $\vec{v}_\mathrm{rot}=\vec{\Omega} \vec{\times} \vec{r}$.
The Keplerian shear \citep{Nakazawa1988} is described by
\begin{equation}
\vec{v}_\mathrm{shear}(x,y,z)=(0,-1.5\Omega_\mathrm{Kep} x,0)
\end{equation}
for a rotating reference frame on a circular orbit with $\Omega_\mathrm{Kep}=(GM_{\sun}/a_\mathrm{hel}^3)^{1/2}$, where $a_\mathrm{hel}=30\, \si{AU}$ is the heliocentric semimajor axis of the cloud orbit.
This distance is typical for the location of the primordial disc where the planetesimals formed \citep{Morbidelli2008,Morbidelli2019}.

\subsection{Choosing the timestep}
\label{sec:timestep}

We chose our simulation timestep to be small enough to accurately calculate particle trajectories, and to consistently detect collisions between particles.
In \rebound, a collision is recorded when two particles are found to be overlapping at the end of a timestep.
The distance a particle moves in a single timestep must be smaller than the target particle size, otherwise an impacting particle could miss the target.
We defined the relative collisional step size parameter as 
\begin{equation}
\label{eqn:rel_step_size}
d_\mathrm{rel}=\mathrm{d}t \cdot v_\mathrm{rel}/d_{f,\, i}
,\end{equation}
where $\mathrm{d}t$ is the timestep, $d_{f,\,i}$ is the diameter of the target simulation particle $i,$ and $v_\mathrm{rel}$ is the relative (collisional) velocity between the target and impactor.
We therefore defined a collision resolution criterion, $d_\mathrm{rel} < 1$, such that when a collision satisfies this inequality we assume that it has been well resolved.
In section \ref{sec:timestep_collisions} we consider the effects of increasing the timestep on mass accretion during gravitational collapse.

We chose the timestep for each simulation from an expected maximum collisional velocity and the initial particle size $d_f=2r_f=2f^*r^*$ such that 
\begin{equation}
\label{eqn:dt}
\mathrm{d}t=\frac{1}{3}\frac{d_f}{v_\mathrm{max}}
,\end{equation}
where we have included a numerical factor of $1/3$ to guarantee the criterion is met, and we have assumed that the maximum collisional velocity during collapse is $v_\mathrm{max}=30\, \si{m\,s^{-1}}$. 
This estimate of the collisional velocity was obtained from \cite{Fraser2017}, supplementary figure 4, but dynamical excitation of particles in the cloud during collapse is highly dependent on the initial cloud mass, with higher mass clouds having higher particle velocities.

To ensure that we achieved accurate collision resolution during gravitational collapse, we verified for each simulation that the relative collisional step size parameter (equation \ref{eqn:rel_step_size}) of all recorded collisions satisfied the collision resolution criterion.
In most cases the criterion was well satisfied, thanks to our conservative choice of safety factor in equation \ref{eqn:dt}.
However, in some simulations of the highest cloud mass, $M_\mathrm{c} = 1.77\e{21}\, \si{kg}$, we found that a moderate number of the collisions exceeded the estimated maximum velocity and so were detected outside the collision criterion.
This meant that by chance we had detected collisions that were faster than expected, which indicated that there could be other collisions that were missed.
Therefore for these particular high cloud mass runs we decreased $\mathrm{d}t$, reducing the numerical factor in equation \ref{eqn:dt} from $1/3$ to $1/6$, and we repeated the simulations.
When the collision criterion was tested for the final dataset, we found that only $0.002\%$ of all recorded collisions had $d_\mathrm{rel}>1$.

\begin{figure}
        \centering
        \includegraphics[]{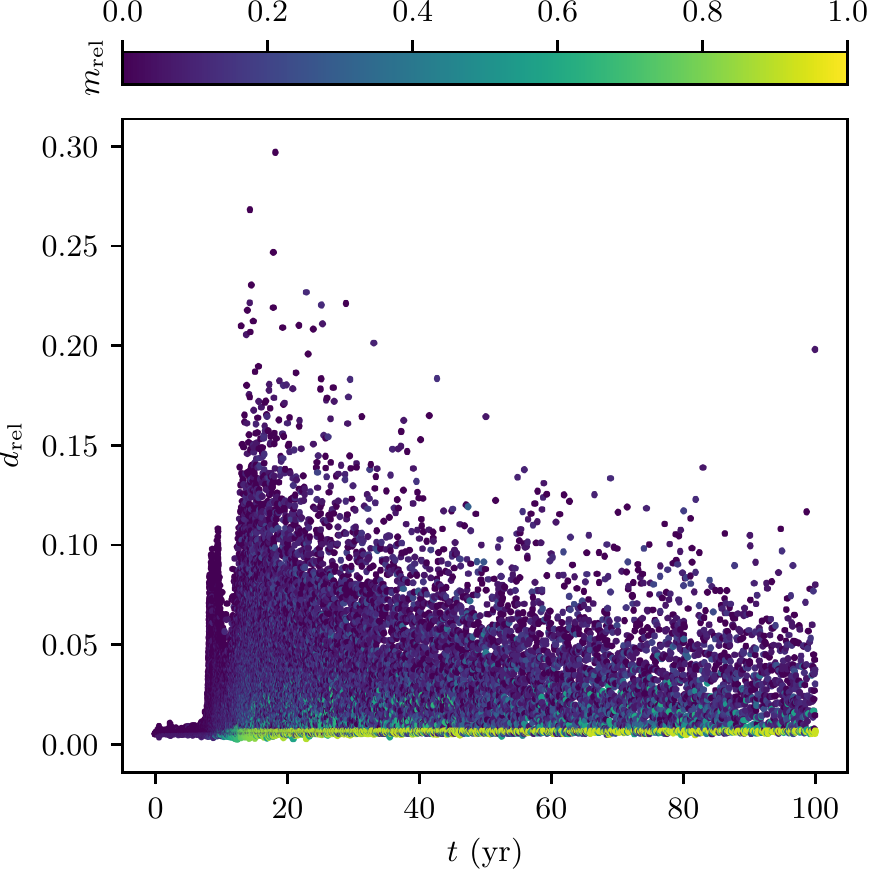}
        \caption{Relative collisional step size (equation \ref{eqn:rel_step_size}) as a function of time for an example simulation ($f^*=10$, $X=0.5$ and $M_\mathrm{c}=6.54\e{19}\, \si{kg}$). Each point represents a collision recorded in this collapsing-cloud simulation. The logarithmic colour scale, $m_\mathrm{rel}=\log (m / m_0)/ \log (M_\mathrm{c} / m_0)$, shows the mass of the primary (most massive) particle that is involved in each collision.
        }
        \label{fig:collision_criterion}
\end{figure}

Even if a small number of collisions are missed, this does not invalidate our results.
With gravitational collapse runaway growth is common, whereby large particles grow at a faster rate than smaller ones \citep{Kokubo1996}, quickly accreting the majority of the cloud mass.
These are the particles that are most likely to harbour binary components, and in this study, only the 100 most massive simulation particles were searched for orbits (section \ref{sec:OrbitSearchAlgorithm}).
The most massive particles were found to always be well within the collision criterion.
Because the criterion depends on particle size, as $d_{f,\,i}$ increased, the criterion became easier to satisfy.
This is demonstrated in figure \ref{fig:collision_criterion}, where we show the relative step size parameter for all collisions in an example gravitational-collapse simulation.
We found that the collisions that fell outside the criterion were always between the smallest particles in the simulation, with $d_{f,\,i}$ close to the initial particle size. 
These fast collisions only occurred after the initial collapse of the cloud, given by the free-fall time
\begin{equation}
t_\mathrm{ff}=\sqrt{\frac{3\pi}{32G\rho_\mathrm{c}}} \label{eqn:free_fall_time}
,\end{equation} 
where $G$ is the gravitational constant, $\rho_\mathrm{c}=3M_\mathrm{c}/4\pi R_\mathrm{c}^3$ is the initial cloud density, and $t_\mathrm{ff}=2.5\e{8}\, \si{s} \simeq 8 \, \si{yr}$ for our initial conditions.
By this stage the bulk of the cloud mass is already located in the most massive particles, therefore we argue that it is inconsequential that there may be a small number of undetected collisions between these small particles.
The large bodies that form in the early phases of gravitational collapse are those that we are most interested in as binary system candidates. 
These bodies dominate the cloud mass fraction and have the greatest gravitational influence, therefore collisions between small bodies will not affect the already accreted larger bodies.

\subsection{Collision resolution}
\label{sec:collision_resolution}

We used the \rebound collision tree module.
This module uses an oct-tree to search for nearest neighbours and records a collision when overlapping particles are detected.
Each simulation recorded a log of all collisions that contained the time and the target, impactor, and merged particle coordinates.
We used the default inelastic merger collision routine, where the colliding particles are replaced by a single particle at the centre of mass of the original two.
The resultant particle conserves total linear momentum, mass, and volume of the constituent particles.
\\

In additional tests, \cite{Nesvorny2010} introduced inelastic bouncing collisions. 
Particles would merge if the impact velocity was lower than a certain threshold, otherwise, they would inelastically bounce with a loss of energy.
This was found to have little effect on their simulations, as the majority of collisions were below the threshold velocity, therefore we also ignored such effects.
To justify this assumption, we evaluated all recorded collisions against the merging collision criterion $v_\mathrm{rel}<v_\mathrm{esc}$ \citep{Leinhardt2012,Mustill2018}, where $v_\mathrm{esc}$ is the escape velocity of colliding particles with masses $m_i$ and $m_j$.
We found that $87\%$ of all recorded collisions satisfied this criterion and should indeed be merging.
It should be noted, however, that in these simulations collisions are detected when the \textit{\textup{inflated}} particle radii overlap. We assessed the increase in collisional velocity if particles continue to accelerate towards each other and collide at their `true' physical radii ($r=(3m/4\pi\rho)^{1/3}$). The percentage of merging collisions drops to $79\%$.
For this estimate, we have assumed that all collisions are head-on. As all collisions have some non-zero impact parameter, our estimate represents an upper limit to the increase in collisional velocity. Regardless, the majority of collisions are still merging.
The most massive particles (high $v_\mathrm{esc}$) always experienced merging collisions, and it was only the low-mass particles at $t>t_\mathrm{ff}$ that could have bounced.

Furthermore, work by \cite{Sugiura2018} shows that collisions between planetesimals (albeit basalt ones) can be merging with no mass loss over an even wider range of impact velocities, depending on the impact angle.
Therefore we assumed that inelastic merging collisions are appropriate when investigating gravitational collapse of the cloud masses used here.
Bouncing collisions should not affect the formation of the most massive particles or any binary systems they may form.

\section{Results}
\label{sec:Results}

\subsection{Searching for bound systems}
\label{sec:OrbitSearchAlgorithm}

After $t=100\, \si{yr,}$ we halted the gravitational-collapse simulation and searched for bound systems of particles.
We searched for orbits amongst a subset of simulation particles using the following criteria.

We are primarily interested in bodies that have undergone significant growth during the simulation; these are the objects that have condensed out of the particle cloud and formed distinct bodies.
First we considered only particles that have had at least one merging collision, and had mass $m\geq2m_0$.
The maximum mass of a binary system we could have detected is then $m_1+m_2\geq4m_0$.
In general, at the end of gravitational collapse, most of the cloud mass was highly concentrated in only a handful of objects, and these particles are the most likely to have a companion.
We also chose to ignore any bound systems in which the satellite would be extremely small and unobservable.
To do this, we applied a mass ratio cut of $m/M>10^{-3}$ to the list of particles to be searched, where the most massive particle in the cloud had mass $M$. 
This is equivalent to an observational magnitude difference of $\Delta \mathrm{m} = 5\, \si{mag}$ for a primary and secondary of the same density and albedo.
Finally, we limited the orbit search to the $N_\mathrm{lim}=100$ most massive particles to ensure a fast and efficient search. 
This was the most significant constraint on the particles included in the orbit search but was appropriate, as explained in section \ref{sec:orbit_detection}.

We took each of these particles to be a potential primary, and we searched all other particles in the subset to see if they were gravitationally bound. 
We first transformed the positions and velocities from the rotating Hill frame into the heliocentric frame. 
Then the orbit of the secondary relative to the primary was found using the \rebound \texttt{calculate\_orbit} function, assuming that this was a bound orbit if the binary eccentricity was $0 \leq e_\mathrm{bin}<1$.
This provides us with a list of bound orbits between pairs of particles.
This list was sorted into independently bound systems of particles using the \texttt{networkx} package \citep{Hagberg2008}, which found all groups of mutually associated orbits. This method automatically accounts for all possible bound orbit architectures, for instance,\ several satellites orbiting a single primary or more complicated nested systems. Furthermore, it clearly shows when more than one bound system is produced in our simulations, that is,\ we detected no connecting orbits between independently bound systems.

At $t=100\, \si{yr,}$ we found that it was common for the collapsed cloud to contain several bound systems, some of which showed a high degree of multiplicity.
We detected a wide variety of bound systems, ranging from simple binary pairs to multiple-body systems of various flavours, for example,\ a single massive primary with a swarm of secondaries or circumbinary systems.
We chose to include all systems of independently bound particles found by the orbit search in our analysis, unlike \cite{Nesvorny2010}, who considered only the single most massive binary produced by each collapsing cloud.
The orbits detected at this stage are instantaneous (osculating) orbital elements, and these systems may not be stable over longer timescales.
It is therefore important to test these bound systems further. 

\subsection{Further dynamical evolution of bound systems}
\label{sec:further_dynamical_evolution}

As in \cite{Nesvorny2010}, we evaluated the stability of these systems by dynamically evolving them for a further $10^{4}\, \si{yr}$.
For each independently bound system of particles detected at $t=100\, \si{yr,}$ a new $N$-body simulation was launched to investigate its dynamical stability.
We integrated the system in the heliocentric frame using the \rebound leapfrog integrator.
During the previous gravitational-collapse simulations the most massive particles grew quickly by runaway growth, and so by $t=100\, \si{yr,}$ their mass accretion was complete.
We therefore removed the inflation factor $f^*$ for these integrations; particles of mass $m$ were initialised with radius $r=(3m/4\pi\rho)^{1/3}$. 
Size inflation was important for boosting the collision rate during gravitational collapse but it prevents us from investigating the dynamics of tight orbits, as particles cannot get close without colliding and merging into a single body.

After the additional integration the orbit search was repeated, and we assessed how the initially bound system had evolved.
In this dataset we included only orbits that had $a_\mathrm{bin}/R_\mathrm{Hill}<0.5$, where $a_\mathrm{bin}$ is the binary orbit semimajor axis, and the Hill radius of the binary primary  of mass $m_1$ is
\begin{equation}
R_\mathrm{Hill}=a_\mathrm{hel}(m_1/3M_{\sun})^{1/3}
\label{eqn:R_Hill_bin}
,\end{equation} 
where $a_\mathrm{hel}\simeq30\, \si{AU}$ is the heliocentric distance of the binary primary.
This was to filter out any poorly bound orbits that we did not expect to survive.

At $t=100\, \si{yr}$,  287 bound systems were detected for the 144 collapse simulations.
Of these systems, $55\%$ were simple binary pairs and the remainder were multiple-body systems.
After $10^{4}\, \si{yr}$ of dynamical evolution we found that either multiplicity $N$ was greatly reduced (most systems evolved to simple binary pairs), or the system was destroyed (either dynamically or collisionally).
We were left with 223 systems: 188 binary systems, and 35 with $N>2$.
Of the systems that were initially binary pairs, $67\%$ had survived, and $63\%$ of the multiple systems had evolved to become $N=2$ binaries.
For the surviving multiple-body systems, 32 had been reduced to triple systems, and 3 systems had $N=4$ bound particles.

\subsection{Properties of binary systems}

For our analysis of the bound systems produced by gravitational collapse we focused on the binary planetesimals. 
After the dynamical evolution, some multiple particle systems ($N>2$) survived in our dataset, but with even further dynamical evolution these systems may be destroyed, evolve into binaries, or may remain as stable $N>2$ systems.
In the following analysis, we only briefly assess the properties of the multiple-body systems by considering the orbit between the two most massive particles in each system.
We made the assumption that the properties of the most massive particles would not be greatly changed by future evolution, provided the system survives.

\subsubsection{Binary masses}
\label{sec:binary_masses}

\begin{figure}
        \centering
        \includegraphics[]{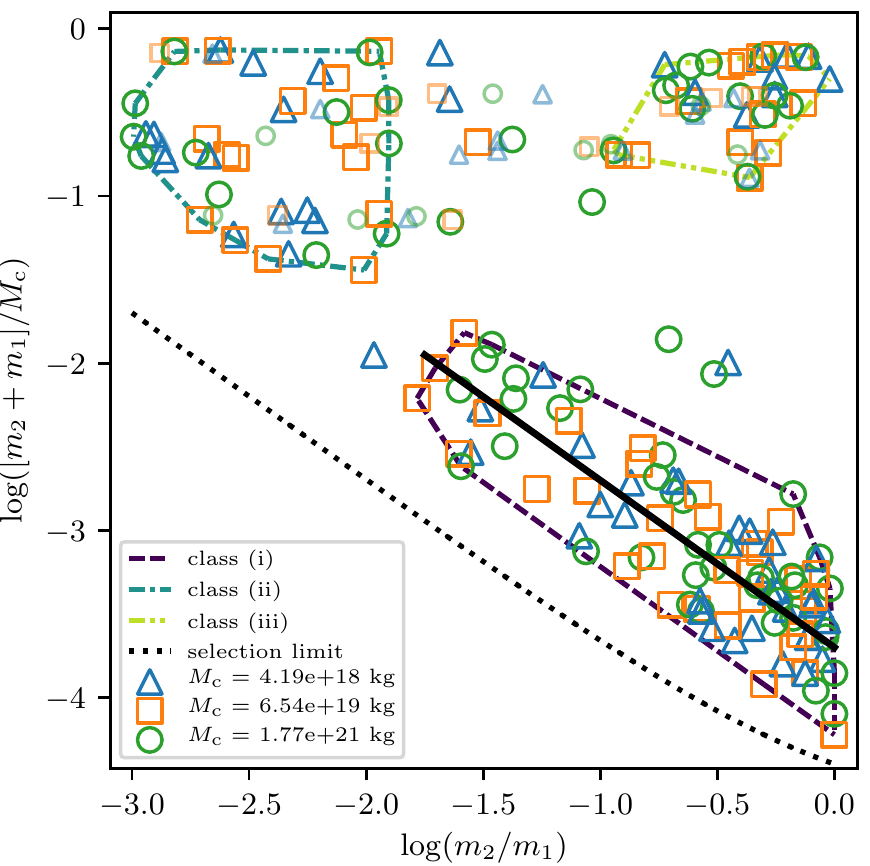}
        \caption{
                Binary system mass normalised by initial cloud mass $(m_2+m_1)/M_\mathrm{c}$, against the secondary-to-primary mass ratio, $m_2/m_1$, plotted in log-log space.
                We display the results for all values of $M_\mathrm{c}$, $X,$ and $f^*$ in our dataset.
                Marker colour and shape denote the three initial cloud masses.
                Dashed coloured lines trace the boundary points of the three binary classes.
                The solid line was drawn to guide the eye to the linear trend of the class (i) particles. 
                The detection limit for our orbit search algorithm is shown as the dotted line.
                We also include data points representative of the $N>2$ bound systems.
                The total mass and mass ratio for the two largest particles in these systems are plotted, represented by the smaller, fainter markers than the binaries.
        }
        \label{fig:binary_log_mass_ratios_mult}
\end{figure}

In figure \ref{fig:binary_log_mass_ratios_mult} we plot the binary mass ratio (secondary mass/primary mass) against the total binary mass normalised by initial cloud mass in log-log space.
A wide range of systems are formed, but there are three distinct populations, which we classify as follows:
\begin{enumerate}
        \item Particles that have undergone minimal mass accretion: $(m_2+m_1)/M_\mathrm{cloud} \sim10^{-3}$, but can have high mass ratios. We refer to these as `atomistic binaries'.
        \item Particles with negligible mass companions: $m_2/m_1 \sim10^{-2.5}$. We call these the `satellite systems'.
        \item Particles that have undergone moderate to high mass accretion and have reasonably sized companions, $(m_2+m_1)/M_\mathrm{cloud} \gtrsim 0.1$ and $m_2/m_1 \gtrsim 0.1$. We classify these as `observable binary' systems.
\end{enumerate}
The orbit search (section \ref{sec:OrbitSearchAlgorithm}) detected a wide range of mass ratios, from $\sim10^{-3}$ (which is the cut-off for our orbit search) up to exactly 1.
The binaries with $m_2/m_1=1$ are composed of particles that have had the exact same number of merging collisions.
Approximately $50\%$ of the binary systems have $m_2/m_1  > 0.1$, which corresponds to a size ratio of $r_2/r_1=(m_2/m_1)^{1/3}=0.46$.
Gravitational collapse does not necessarily have a preference to form similar size binaries.
The binary system masses span a range from $6\e{-5}\, \si{M_\mathrm{c}}$ up to $0.75\, \si{M_\mathrm{c}}$.

To test for the presence of these classes, we performed a Scikit-learn DBSCAN cluster search \citep{Ester1996,Pedregosa2011} on the binary dataset features, $\log ([m_2+m_1]/M_\mathrm{c})$ and $\log (m_2/m_1)$.
These features were first standardised, that is,\ the mean was removed and the data scaled to unit variance.
DBSCAN then finds clusters by grouping particles that have a minimum number of neighbours within a distance parameter; in this search we used the default values 5 and 0.5 for these parameters, respectively.
The algorithm found the three proposed classes with a small number of outliers.
It is clear that the distribution of the data is invariant with different cloud masses and that the class (i) atomistic binaries follow a linear trend; a line is drawn on figure \ref{fig:binary_log_mass_ratios_mult} to guide the eye.
It may be possible that this trend extends to lower mass ratios for a subset of the satellite systems, but for now we consider only the atomistic binaries.
This trend suggests that these objects follow a power-law relation between the binary system mass (i.e.\ the mass accretion efficiency of gravitational collapse) and the binary mass ratio of the form
\begin{equation}
        \frac{m_2}{m_1}=  10^{-b/s} \alpha^{1/s} \label{eqn:linear_trend}
,\end{equation}
where we define the mass accretion efficiency as $\alpha = (m_2 + m_1)/M_\mathrm{c}$. 
This linear trend (in log-log space) can be approximated by a gradient and intercept of $s=-1.0$ and $b=-3.7$ respectively, such that $m_2/m_1 \propto \alpha^{-1}$.

This either indicates an underlying physical relation between mass accretion and mass ratio for the atomistic binaries formed during gravitational collapse, or that there are unknown biases in our simulation method.
To investigate this, we considered the selection limit of the orbit search algorithm.
As described in section \ref{sec:OrbitSearchAlgorithm}, we required that both components had undergone at least one merging collision.
Particles must have $m\geq 2\e{-5}M_{\mathrm{c}}$ to be searched, therefore the lowest detectable binary system mass is $4\e{-5}M_{\mathrm{c}}$.
As we also imposed $m_2/m_1\geq10^{-3}$,  there is a limit to the lowest secondary mass that our algorithm will detect for a given primary mass, which we show as a dotted line in figure \ref{fig:binary_log_mass_ratios_mult}.
However, all binaries in our dataset are above this selection limit. 
It is also possible that this trend could be a numerical artefact of the $N$-body method used in this study.
Higher resolution simulations with more particles are required in order to verify this trend.
On the other hand, if this relationship is truly physical, it implies that gravitational collapse produces some binaries where the mass ratio depends primarily on the collapse accretion efficiency.
This is described by equation \ref{eqn:linear_trend}, with exact coefficients that depend on the (currently approximated) slope of the trend line. 
Thus this trend may provide an observationally detectable signature indicating cloud collapse as a real formation route for binary systems.
\\

We then evaluated the $N>2$ systems, which make up $16\%$ of our dataset.
In figure \ref{fig:binary_log_mass_ratios_mult} we include data points that represent what a multiple-body system would look like as a binary if we considered only the properties of the two most massive particles in the system.
We made this assumption as most multiples in our dataset were in a circumbinary configuration, where a tight inner binary is orbited by a more distant lower mass satellite.
Unlike the binary pairs, which generally had high mass ratios, the $N>2$ systems all had relatively low mass ratios. The median mass ratio for the two most massive particles in these systems was $m_2/m_1=0.03$, therefore we would expect only minor gravitational perturbations from the third (or fourth) body.
It is likely that such a system could initially be discovered as a binary, as the inner most satellite may be unresolved, or that a faint, distant satellite is below the detection limit.
With further observation, it could be revealed that what appears to be a binary is actually a multiple system, for example 1999 TC$_{36}$ Lempo \citep{Benecchi2010}.
In figure \ref{fig:binary_log_mass_ratios_mult} the $N>2$ systems fall into the mass range of observable binaries and satellite systems, that is,\ the main orbit of a multiple-body system contains a reasonable fraction of the collapsing cloud mass.
The mass ratios between the two largest bodies, however, span the full range of possible values.
Multiple-body systems appear to bridge the gap between the two classes, adding to the population of outlier points.
We repeated the DBSCAN cluster search for both binaries and multiples using the same parameters as before, and found that classes (ii) and (iii) were detected as a single large cluster.
Hence it is possible that observationally, only these two groupings would be detected, although we note that in figure \ref{fig:binary_log_mass_ratios_mult} the high-multiplicity systems more frequently have either high or low $m_2/m_1$.
These systems would likely fall into the observable binary and satellite system categories.
Moreover, we did not consider how an $N>2$ system might evolve for timescales longer than $10^{4}\, \si{yr}$.
Further integration is required to investigate the stability of such systems over longer timescales to the present day.
Collisions between components may change the mass ratios, a component may be ejected, or the system could be destroyed altogether.
\\

\begin{figure}
        \centering
        \includegraphics[trim={0 0 0 0.03cm}]{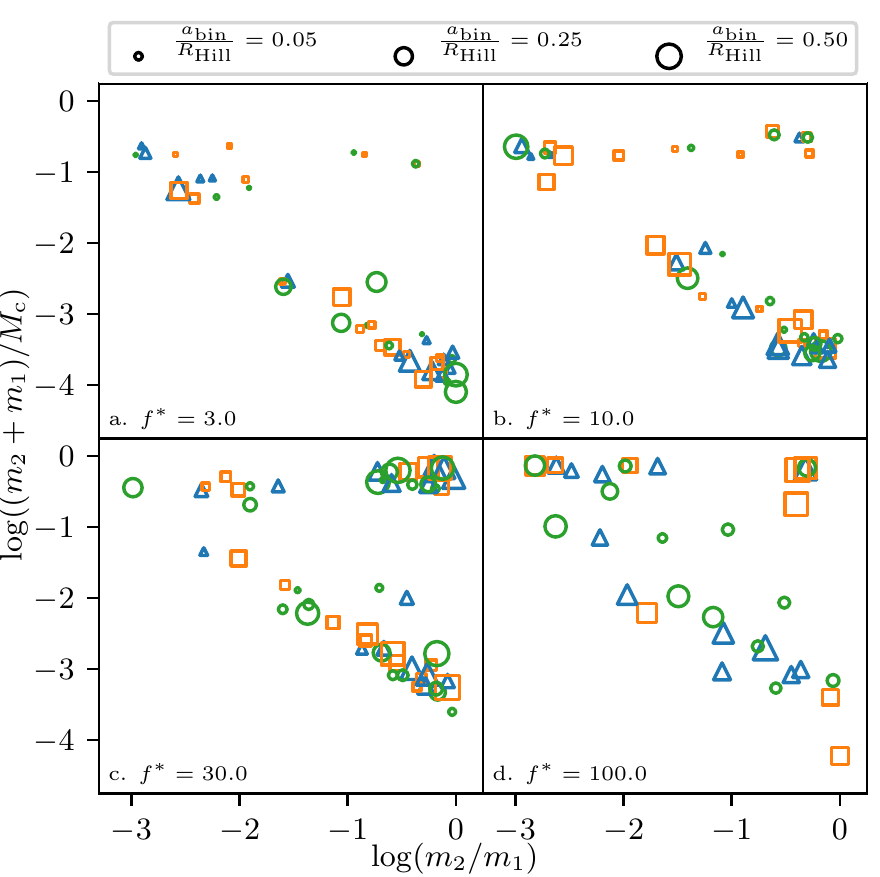}
        \caption{Breakdown of figure \ref{fig:binary_log_mass_ratios_mult} (binaries only) by particle inflation factor $f^*$. 
                Marker colours and shapes are the same as before, and marker size scales with relative binary separation, $a_\mathrm{bin}/R_\mathrm{Hill}$. 
                Panels a, b, c, and d show the binaries produced by simulations with $f^*=3,\ 10,\ 30, \text{and}\ 100,$ respectively.
        }
        \label{fig:binary_mass_ratios_by_f}
\end{figure}

The data (binary orbits only) separated by the different values of particle inflation factor $f^*$ are shown in figure \ref{fig:binary_mass_ratios_by_f}.
Marker size indicates the separation of binary components relative to the Hill radius of the primary (equation \ref{eqn:R_Hill_bin}).
We see that simulations with $f^*=30$ produced the most observable binary systems.
For low values of $f^*$ , the particle collisional cross section is small and the reduced number of high-mass systems is explained by the low collision rate between particles.
This is supported by the systems that are detected for $f^*=3$ and $10$ mostly belonging to the low-mass atomistic binary group.
As $f^*\rightarrow30$, collision rate and mass accretion increases and more high-mass systems with $m_1+m_2>0.1M_\mathrm{c}$ are produced.

\begin{figure}
        \centering
        \includegraphics[]{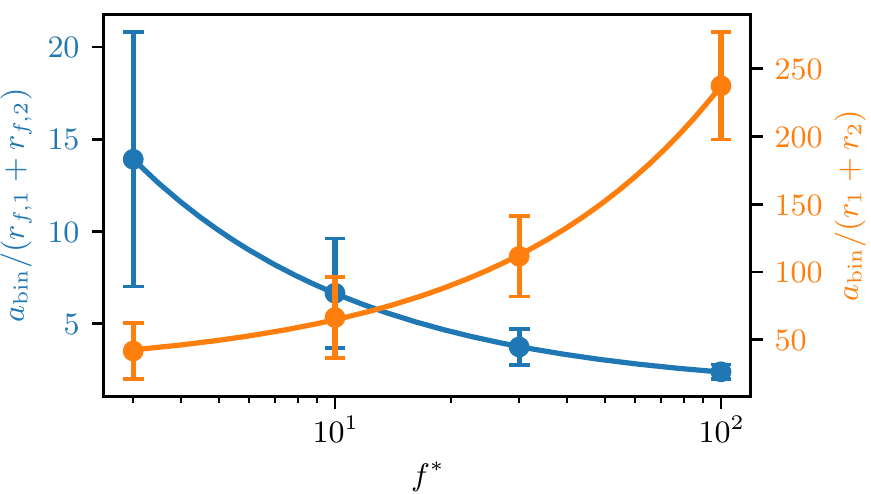}
        \caption{
                Median binary separation relative to the particle size as a function of particle inflation factor $f^*$ used in the simulation that formed the binary. 
                        The left-hand axis (blue) shows separation relative to the inflated particle radii, and the right-hand axis (orange) is relative to the uninflated physical particle radii.
                        The error bars indicate the standard error for each set of data points from simulations with a given $f^*$. 
                Power-law fits are drawn to guide the eye.
        }
        \label{fig:binary_rel_sep}
\end{figure}

When $f^*=100,$ the number of high-mass binaries drops.
With high $f^*$ , we would expect an increased collision rate, but instead we hit the limit where these extremely large inflated particles cannot form a tight binary without colliding and destroying the system.
To show this, we considered the binary separation relative to the particle size, both with and without particle size inflation. In figure \ref{fig:binary_rel_sep} we plot the median relative separation of all binaries formed by simulations with a particular value of $f^*$.
When the increase in particle size due to $f^*$ is accounted for, the relative separation is given by $a_\mathrm{bin}/(r_{f,\,1}+r_{f,\,2})$. Simulations with larger $f^*$ generally produce binaries that are tight relative to the inflated particle size (figure \ref{fig:binary_rel_sep}, blue line). This makes such systems more vulnerable to destruction by mutual collision. When the relative separation is recalculated, but instead using the physical particle radii, $a_\mathrm{bin}/(r_{1}+r_{2})$, the trend is now that relative separation increases for simulations run with higher values of $f^*$ (figure \ref{fig:binary_rel_sep}, orange line). This means that in addition to reducing the number of binaries formed (due to mutual collisions between components), simulations with higher values of $f^*$ are biased against producing binaries with tight orbits.

\subsubsection{Binary sizes}
\label{sec:binary_sizes}

\begin{figure}
        \centering
        \includegraphics[]{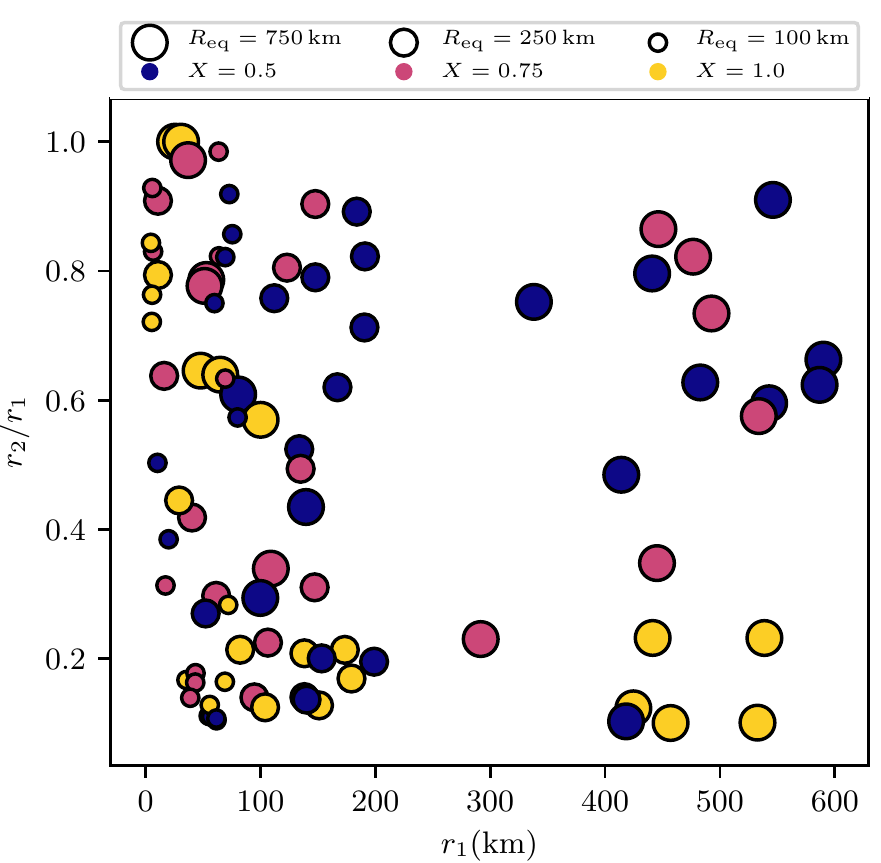}
        \caption{Binary size ratio against primary size from our simulations, shown in a similar manner to figure 2 of \cite{Nesvorny2010}.
                Marker size indicates the cloud mass (given as an equivalent size $R_\mathrm{eq}$), and colour gives the cloud rotation. As in the original, we show only the results for $f^*=3,\,10,\,30$.
        }
        \label{fig:binary_size_ratios}
\end{figure}

To compare our data with the binary systems of \cite{Nesvorny2010} we recreated their figure 2, a plot of the binary secondary/primary size  $r_2/r_1=(m_2/m_1)^{1/3}$ against the (uninflated) primary radius $r_1$ (figure \ref{fig:binary_size_ratios}).
For consistency with their work, we only considered binaries produced by clouds with moderate values of inflation factor, $f^*=3$, 10, 30.
However, we found that including the $f^*=100$ binary systems did little to change the overall distribution.

Comparing the figures, we note that our implementation of gravitational collapse produced more massive binaries.
\cite{Nesvorny2010} produced a maximum primary radius of $r_1\simeq400\, \si{km}$ for the highest cloud mass, which corresponds to a fraction $(400/750)^{3}=0.15$ of the initial cloud mass.
In contrast, the largest particle in figure \ref{fig:binary_size_ratios} is approximately $(600/750)^3=0.5$ of the initial cloud mass.
As shown in figure \ref{fig:binary_log_mass_ratios_mult}, we found that gravitational collapse generally accretes a large fraction (up to $75\%$) of the cloud mass into a binary system, and that this process is invariant with $M_\mathrm{c}$.
The cloud masses span three orders of magnitude, and if the binaries all accreted approximately the same mass fraction, the particle sizes will also cover a wide range.
This explains why we have a less uniform distribution in $r_1$ than \cite{Nesvorny2010}.  
This increased mass accretion efficiency in our implementation of gravitational collapse may be a result of enhanced collision resolution.
As described in section \ref{sec:collision_resolution}, we tailored the timestep for each cloud to ensure collisions were accurately detected.
\cite{Nesvorny2010} used a longer, fixed timestep of $0.3\, \si{days}$ for all cloud masses, but with a collision detection method that extrapolates particle trajectories to allow for longer timesteps \citep{Richardson2000}. 
We investigate this further in section \ref{sec:timestep_collisions}.

Furthermore, we found a wider spread in binary size ratio, but this is to be expected as we have included all binaries found by the deep orbit search. 
In figure \ref{fig:binary_size_ratios} we reproduce the trend of \cite{Nesvorny2010} that most large, equal-sized binaries were produced by clouds with rotation $X<1$.
In contrast, we found that some $X=1$ clouds were able to produce large systems, but these are all low size ratio satellite systems.
At the other extreme, we detected binaries from $X=1$ clouds with small primaries and an equal-size secondary: these are the atomistic binaries.
Although the initial angular momentum of the cloud would be most efficiently `spent' by forming a large similar-size binary, in the $X=1$ case we reach the limit where cloud collapse is resisted by the cloud rotation. Particles are more readily dispersed and ejected by the Keplerian shear, leading to a loss of angular momentum. Therefore satellite systems and atomistic binaries are preferentially formed in $X=1$ simulations.
When we take into account the larger particles and wider variety of systems that we included in this study, our implementation of gravitational collapse produced a subset of systems of appreciable mass and equal-sized components, similar to \cite{Nesvorny2010}.

\subsubsection{Binary magnitudes}
\label{sec:binary_magnitudes}

We then compared our binary systems with all the observed TNBs from the \cite{GrundyWeb} database (accessed 2 December 2019, 102 objects).
Within this dataset some systems are classified as `special cases', for example the known triple system Lempo \citep{Benecchi2010}, the contact or  nearly contact binary 2001 QG$_{298}$ \citep{Sheppard2004,Lacerda2011}, and some are placeholders for known binaries with incomplete orbits.
We excluded the unusual triple and contact systems.
The dynamically complex Pluto and Haumea systems were not included in this dataset, but we did include dwarf planet systems such as Eris and Makemake, plus the `nearly certain' dwarf planet candidates: 2007 OR$_{10}$, Quaoar, Orcus, and Salacia (\cite{BrownWeb}, accessed 3 December 2019).

\begin{figure*}
        \sidecaption
        \includegraphics[]{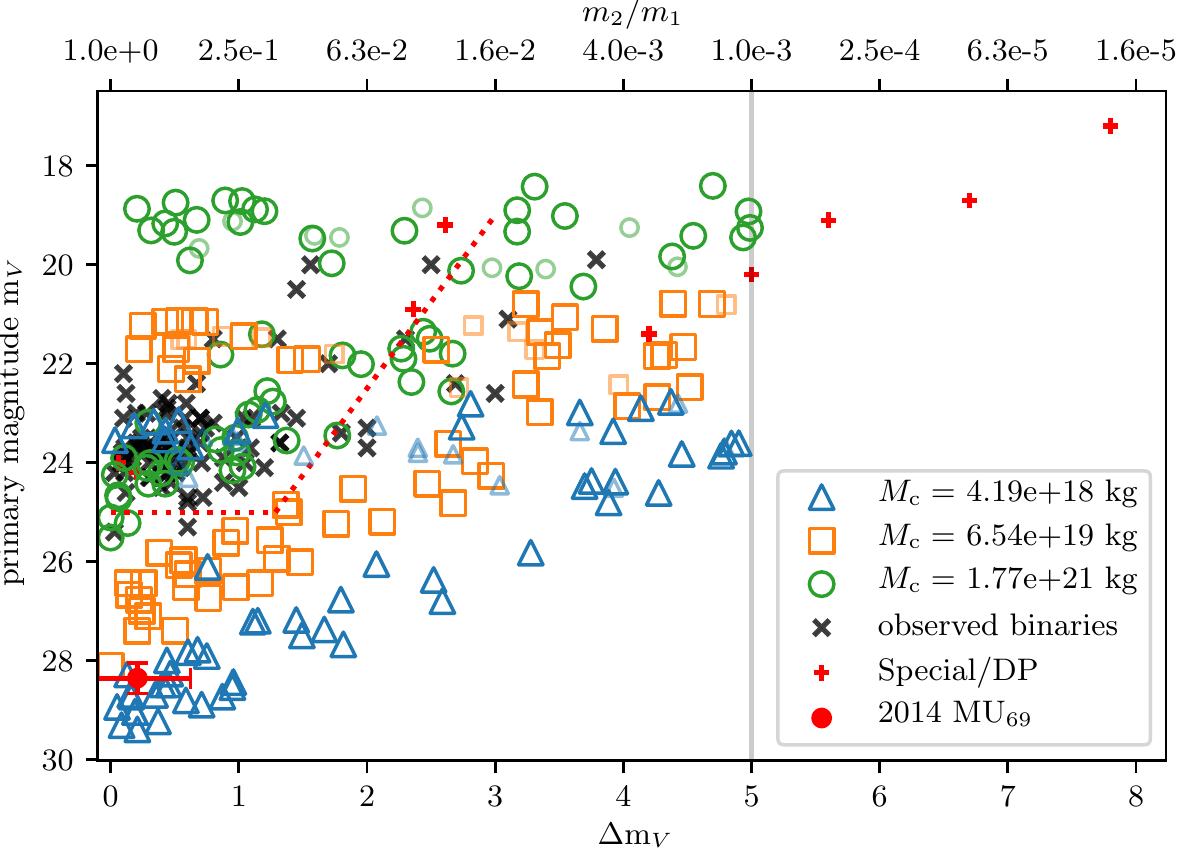}
        \caption{Primary $V$ -band magnitude against magnitude difference between primary and secondary for binaries produced by gravitational-collapse simulations.
                Particle masses were converted into a spherical radius (assuming uniform density), which was then converted into a magnitude using a fixed albedo and distance.
                As with figure \ref{fig:binary_log_mass_ratios_mult}, marker colour and shape indicate the initial cloud mass, and any $N>2$ systems are represented by smaller, fainter markers.
                The mass ratio for a given magnitude difference is shown on the upper $x$ -axis. The vertical line indicates the mass ratio cut-off of $m_2/m_1\geq10^{-3}$ in the orbit-search algorithm. 
                The observed TNBs from \cite{GrundyWeb} are shown as black $\text{crosses}$. We represent `special cases' and dwarf planets as red $\text{plusses}$. 
                The red circle with error bars represents how Arrokoth (2014 MU$_{69}$) would appear if its components could be separately resolved.
                An approximate empirical detection limit \citep{Noll2008} and a lower magnitude limit of 25 are shown as dotted red lines. 
        }
        \label{fig:binary_magnitudes}
\end{figure*}

Similar to figure 3 of \cite{Nesvorny2010}, we evaluated the magnitude of the binary primary as a function of the primary-secondary magnitude difference for the simulated binaries and observed TNBs (figure \ref{fig:binary_magnitudes}).
We used the uninflated radii\footnote{Similar to sections \ref{sec:collision_resolution} and \ref{sec:further_dynamical_evolution}, the uninflated particle radius was calculated from mass $m$ and density $\rho.$} for the primary and secondary particles, $r_1$ and $r_2$, to calculate an observed $V$ -band magnitude for each binary system (Appendix \ref{app:magnitudes}).
Magnitudes were calculated at a heliocentric distance of $R=44\, \si{AU}$ to compare with the observed TNBs, most of which are located in the classical belt.
We highlight that the gravitational-collapse simulations were set up at $30\, \si{AU}$, therefore we have to assume that either the binaries have migrated from their formation region, or that streaming instability and gravitational collapse are not strongly dependent on heliocentric distance \citep{Yang2017}.
A geometric albedo of $p_V=0.15$ was used for both components, which is the observed median albedo for cold classical objects \citep{Lacerda2014}, whereas \cite{Nesvorny2010} used $p_V=0.08$.
We also included data points representing the $N>2$ particle systems formed by gravitational collapse, as in figure \ref{fig:binary_log_mass_ratios_mult} above.
These data points were added to show what a multiple system may look like if it were to be initially discovered as a binary.
Furthermore, in figure \ref{fig:binary_magnitudes} we included a line representing an approximate empirical detection limit \citep[see][figure 3]{Noll2008}, and a line at magnitude $\mathrm{m}_V = 25$ to represent a rough observational brightness limit.

We also added a data point that is representative of the contact binary Arrokoth, with magnitudes calculated using the same method as the binaries above.
The volumes of the primary and secondary components are $V_1=1400\pm600\, \si{km^3}$ and $V_2=1050\pm400\, \si{km^3}$ , respectively \citep{Stern2019}.
To compare with the simulated binaries, we assumed that the components are spherical and therefore have radii $r_1=6.9\pm1.0\, \si{km}$, $r_2=6.3\pm0.8\, \si{km}$.
We considered the hypothetical case where these components are separately resolved such that we can calculate a primary magnitude and a magnitude difference.
We assumed the same observational parameters for distance and albedo as used above.
Compared to the true distance $R=44.6\, \si{AU}$ and albedo $p_V=0.165$ \citep{Stern2019}, the difference in magnitude is minimal.
\\

Figure \ref{fig:binary_magnitudes} shows that each cloud mass produced a similar distribution of binary systems, composed of a linear `fan' and a `clump'.
For each cloud mass these different size regimes are a result of the additional systems found by the deep orbit search (section \ref{sec:OrbitSearchAlgorithm}).
The clump is composed of the observable binaries, which have high system mass and mass ratios.
The linear fan is composed of atomistic binaries and satellite systems.
We compare this to figure \ref{fig:binary_log_mass_ratios_mult}, where these classes were defined and the linear trend of the atomistic binaries was first noticed.
It is interesting that the observed TNBs follow a similar linear trend as the fan, but we would expect that in this case it is due to detection limits and observational biases.
In particular, high cloud mass binaries in the fan lie in a similar parameter space as the observed TNBs with large $\Delta \mathrm{m}_V$ (figure \ref{fig:binary_magnitudes}).
We emphasise again that our binary data points are above the orbit-search selection limit (figure \ref{fig:binary_log_mass_ratios_mult}), and this implies that there is either a physical mechanism or simulation bias that causes this trend.
Figure \ref{fig:binary_log_mass_ratios_mult} also demonstrated that the distribution of binary mass properties is invariant when normalised by cloud mass.
Figure \ref{fig:binary_magnitudes} shows that each cloud mass follows a similar distribution, but with an offset in brightness.
More massive clouds produce more massive, brighter binaries.

In figure \ref{fig:binary_magnitudes} we find that the observable binaries from the low cloud mass and the atomistic binaries from the high cloud mass are the best match to the main cluster of observed TNBs with $\mathrm{m}_V\sim22$ -- 25 and $\Delta \mathrm{m}_V\sim0$ -- $1\, \si{mag}$.
In terms of having high system mass and high mass ratios, our observable binaries are comparable to the binaries detected by \cite{Nesvorny2010}, who selected the single most massive binary from each cloud.
For the atomistic binary systems, although they are a small fraction of the initial cloud mass, $M_\mathrm{c}=1.77\e{21}\, \si{kg}$ is orders of magnitude more massive than the other clouds.
Therefore these two binary classes from different clouds occupy a similar size and brightness range.

In contrast, \cite{Nesvorny2010} found that the intermediate cloud mass best replicated the observed TNBs.
This discrepancy arises firstly because the binaries produced in our gravitational-collapse simulations are generally larger (figure \ref{fig:binary_size_ratios}), and we have used a higher albedo.
This means that our binaries are brighter, which shifts the distribution upwards in figure \ref{fig:binary_magnitudes}, pushing our intermediate cloud mass observable binaries away from the region of the observed TNBs.
Secondly, we have investigated all bound systems produced by gravitational collapse, whereas \cite{Nesvorny2010} did not consider atomistic binaries.

We emphasise that there are many tunable parameters in this analysis, such as heliocentric distance, albedo, and density.
For example, if the material density was decreased to $250\, \si{kg \, m^{-3}}$ objects would appear $\sim1$ mag brighter.
On the other hand, we did not take possible increases in density caused by compaction from collisions or differentiation due to self-gravity into account.
Furthermore, we have shown that the distribution of relative system mass and mass ratio is invariant with $M_\mathrm{c}$, therefore we expect that increasing or decreasing the initial cloud mass would shift the magnitude distribution to brighter or fainter values.
Depending on the mass distribution of pebble clouds formed by streaming instability \citep{Li2019,Nesvorny2019}, clouds might be found whose binary masses match the observed magnitudes better. 

Interestingly, figure \ref{fig:binary_magnitudes} shows a distinct clump of simulated high cloud mass binaries that have no observational counterpart amongst the TNBs.
These are objects that would have magnitudes comparable to the dwarf planets, with a similar size companion, and should be easily detected.
If these objects are not observed, perhaps these binaries are preferentially destroyed by some mechanism.
For example, tidal evolution may shrink the orbit until the components collide and merge into a single object.
An additional factor is that it is relatively rare for the green points in figure \ref{fig:binary_magnitudes} to lie in the clump, compared to the large number of binaries in the fan (12 out of 66 binaries are in the clump for the high-mass cloud dataset).
A more likely explanation, however, is that such high cloud masses simply never existed, and therefore these objects could not be produced by gravitational collapse.
First of all, most cold classicals are observed to have a characteristic diameter $d\simeq140\, \si{km}$, as this marks the break from the steep size distribution of larger objects \citep{Fraser2014}.
Therefore high-mass clouds produce objects that are too large to match observations.
Secondly, the streaming instability simulations of \citet{Nesvorny2019} produce a maximum cloud mass of $\sim10^{21}\, \si{kg}$ (depending on disc parameters, Appendix \ref{app:streaming_instability}).
From these estimates, we see that the highest cloud mass is probably too high to be easily formed by streaming instability.
Therefore it seems unlikely that the high system mass, high-mass ratio clump of $M_\mathrm{c}=1.77\e{21}\, \si{kg}$ binaries shown in figure \ref{fig:binary_magnitudes} can be formed.

We note that there is also the (rather disappointing) possibility that the observable binary feature may only arise due to our limited simulation method, in particular, the assumptions of inelastic merging collisions (section \ref{sec:collision_resolution}) and particle size inflation (section \ref{sec:sim_limitations}). We have justified these assumptions above, but ultimately, more accurate models of gravitational collapse (e.g.\ higher numerical resolution) are required to definitively confirm that the outcomes of these assumptions are indeed accurate reflections of reality.

\subsubsection{Ratio of binary to single planetesimals}
\label{sec:observable_binary_single_ratio}

\begin{figure}
        \centering
        \includegraphics[]{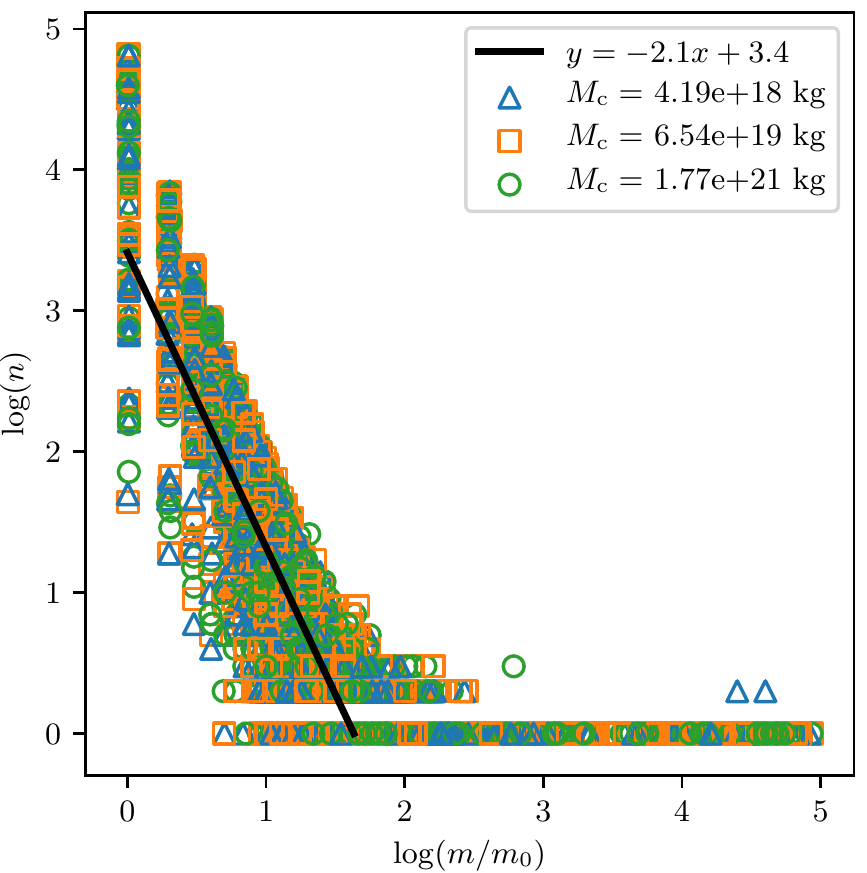}
        \caption{Mass distribution of all particles in the simulation box at $t=100\, \si{yr}$. The number of particles $n$ of mass $m$, in units of the initial particle mass $m_0=M_\mathrm{c}/N_\mathrm{p}$, are shown in log-log space.
                Marker colour and shape denote the three cloud masses. 
                A linear regression on all data points with $\log(m/m_0)<2$ is shown to guide the eye to the power-law distribution.
        }
        \label{fig:particle_mass_dist}
\end{figure}

We also considered the number of particles produced by gravitational collapse, which may be singles or in a bound system, that would be above the observational limit of $\mathrm{m}_V =25$.
In figure \ref{fig:binary_magnitudes} this brightness is equivalent to a simulation particle mass of $m_{25}=1.4\e{17}\, \si{kg}$.
The distribution of particle masses produced by gravitational collapse is a steep power-law, with a particle number that drops sharply with increasing mass.
For each cloud in our dataset, we plotted the mass distribution of all particles in the simulation box at $t=100\, \si{yr}$, as shown in figure \ref{fig:particle_mass_dist}.
By assessing $m$ in units of the initial particle mass, $m_0$, we see that all three cloud masses follow similar distributions.
At low particle mass $\log(m/m_0)<2$, the mass distribution generally follows a power-law of the form $\mathrm{d}n \propto m^q \mathrm{d}m,$ where $q=-2.1$, with roughly two orders of magnitude scatter about that power-law.
At higher masses the distribution flattens off; each simulation produces only a small number of particles that contain a high fraction of the initial cloud mass.

For the low, intermediate, and high cloud mass simulations, the mean number of single particles with a brightness of $25\, \si{mag}$ or greater is $n_\mathrm{single}(m \geq m_{25})=0.3, 1.4, \text{and }147.2,$ respectively. 
When we consider our entire dataset (binaries and multiples), the 144 gravitational-collapse simulations produced 160 bound systems with $m_\mathrm{sys} \geq m_{25}$.
The low-, intermediate-, and high-mass clouds produced on average $n_\mathrm{bound}(m_\mathrm{sys} \geq m_{25})=0.8,1.0,\text{and }1.5$ observable bound particle systems per cloud. 
Therefore, when we compare $n_\mathrm{single}$ to $n_\mathrm{bound}$, only the low cloud masses are compatible with forming nearly all observable planetesimals as binaries, as proposed by \cite{Fraser2017}, with $\sim2.6$ binaries produced for every single planetesimal.
This is further evidence against the gravitational collapse of high-mass clouds forming TNBs, and supports an upper mass limit on clumps formed by streaming instability.

\subsubsection{Binary orbital elements}
\label{sec:binary_orbits}

\begin{figure}
        \centering
        \includegraphics[]{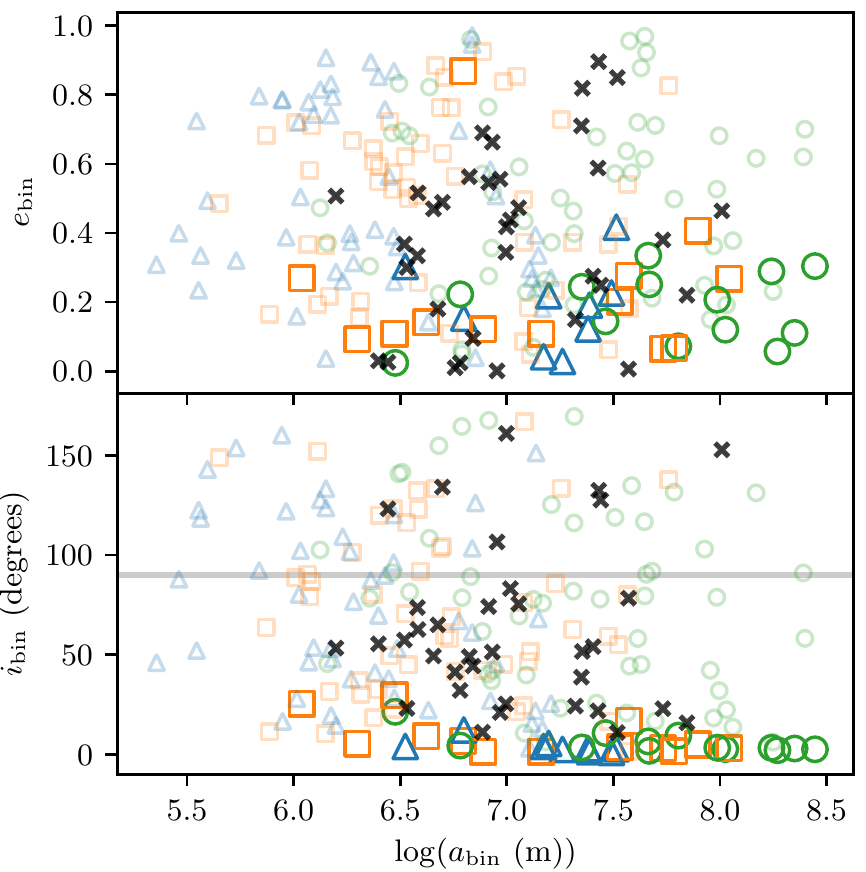}
        \caption{Binary orbital parameters for the simulated binaries compared to the 35 observed TNBs (black $\text{crosse}$s) with full orbit solutions \citep[Table 19]{Grundy2019}.
                Again, marker shape and colour represent the cloud masses (figure \ref{fig:binary_log_mass_ratios_mult}), but here we use larger bold points to highlight systems with $m_2/m_1>0.1$ and $(m_2+m_1)/M_\mathrm{c}>0.1$, i.e.\ the observable binary systems.
                The smaller faint points represent all other systems in the dataset, i.e.\ the satellite systems and atomistic binaries.
        }
        \label{fig:binary_orbits}
\end{figure}

The distribution of binary orbital elements produced by gravitational collapse: semimajor axis $a_\mathrm{bin}$, eccentricity $e_\mathrm{bin}$ , and inclination $i_\mathrm{bin}$ are shown in figure \ref{fig:binary_orbits}.
These are compared to the 35 observed TNBs for which there are full orbital solutions \citep{Grundy2019}.
For the observed TNBs the reported inclination is the angle between the mutual binary orbit and its heliocentric orbit.
The simulated binary inclinations are measured relative to the $xy$ plane of the rotating reference frame, that is,\ relative to the heliocentric orbital plane of the pebble cloud. 
We see that each cloud produces a wide spread of binary orbits, spanning the full range of eccentricity and inclination.
There is a trend for larger binary semimajor axis with increasing cloud mass, which is primarily caused by the longer mean free path of particles in more massive clouds.
We have highlighted the orbital properties of the observable binaries with bold markers.
These objects are analogous to the binary systems found by \cite{Nesvorny2010}; they generally have low prograde inclinations and low to moderate eccentricities.
\\

\begin{figure}
        \centering
        \includegraphics[]{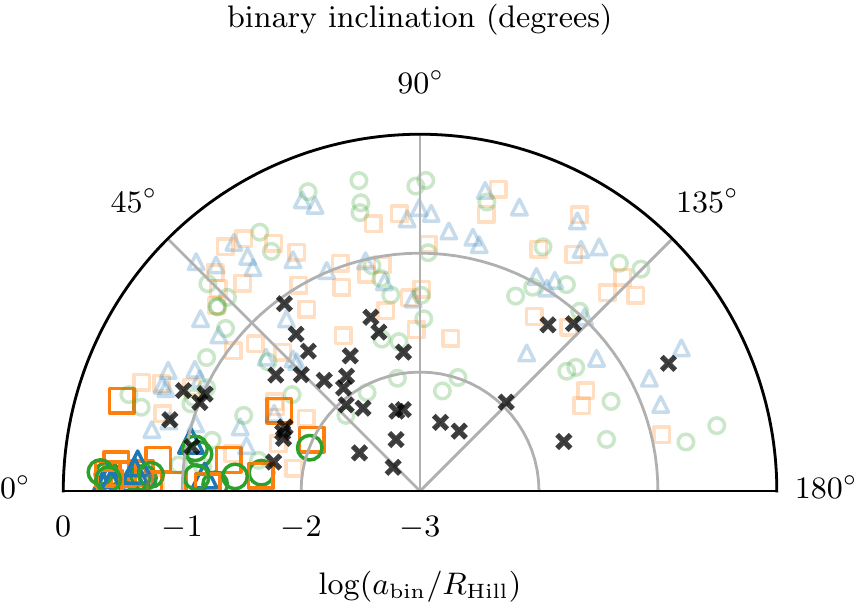}
        \caption{Binary mutual orbit inclination against semimajor axis relative to the Hill radius, recreated from figure 1 of \citet{Grundy2019}. 
                We overplot the binary systems from our simulations using the same marker shapes and colours as in figure \ref{fig:binary_log_mass_ratios_mult} to represent the cloud mass.
                As in figure \ref{fig:binary_orbits}, the larger bold points highlight the observable binary systems and the observed TNBs are black $\text{crosses}$. 
        }
        \label{fig:binary_inclination_polar}
\end{figure}

We then reproduced figure 1 of \cite{Grundy2019}, a polar plot of binary mutual inclination on the azimuthal axis against the relative separation on the radial axis for the 35 observed TNBs with full orbit solutions (figure \ref{fig:binary_inclination_polar}).
For comparison we have included our simulated binary systems, where $R_\mathrm{Hill}$ was calculated at $30\, \si{AU}$, which is the formation distance of binaries through gravitational collapse in this study.
Gravitational collapse does not reproduce the tightest TNBs, that is,\ those with $\log (a_\mathrm{bin} / R_\mathrm{Hill})<-2$.
This could once more be due to the biases in the simulation caused by particle inflation $f^*$ preventing close systems from forming (section \ref{sec:binary_masses}).
Alternatively, it may be that that the systems produced by gravitational collapse evolve onto tighter orbits after formation.
As pointed out by \cite{Nesvorny2019}, Kozai cycles and tidal friction (KCTF), which leads to the tightening and circularisation of orbits \citep{Porter2012}, can be particularly strong for wider binaries above a critical separation $a_\mathrm{bin}/R_\mathrm{Hill}>0.1$
at inclinations of $i_\mathrm{bin}\sim90\degr$.
This could explain what would happen to the excess of wide high-inclination objects in our dataset, compared to the dearth of observed TNBs with such orbits in figure \ref{fig:binary_inclination_polar}, over the long period of time after their formation.

In the simulated binary dataset $29\%$ of systems have retrograde orbits.
In comparison, the observed TNBs have $20\%$ retrograde orbits \citep{Grundy2019}, although we note that the majority of the simulated binaries lie outside the observational limits in figure \ref{fig:binary_magnitudes}.
When we consider only the observable binaries, highlighted in figures \ref{fig:binary_orbits} and \ref{fig:binary_inclination_polar}, these systems have a $100\%$ prograde low-inclination distribution.
Assuming they formed through gravitational collapse, the observed TNB inclination distribution is a combination of the inclination of pebble clouds formed by streaming instability and the subsequent $\sim4\, \si{Gyr}$ of Solar System evolution.
The simulated binary inclinations are given relative to the $xy$ plane of the Hill frame, which is also the heliocentric orbital plane of the rotating pebble cloud (which we have assumed always rotates about the $\hat{\vec{z}}$ -axis). 
This means that we have not taken into account that $\sim20\%$ of pebble clouds generated by streaming instability rotate retrograde, and so would any binaries (that we would classify as observable) that formed from them  \citep{Nesvorny2019}.
This is because the findings of \citet{Nesvorny2010} showed that the inclination of binaries produced by gravitational collapse was low and prograde relative to the collapsing cloud.
This observation was used by \cite{Nesvorny2019} to directly equate the obliquity of a pebble cloud formed by streaming instability to the inclination of the binary system that is formed by gravitational collapse of that cloud.
To directly compare our simulated binary orbits to the observed TNBs in figure \ref{fig:binary_inclination_polar} we would need to apply the streaming instability cloud inclination distribution and the effects of post formation evolution.
This would include not only KCTF, but also the collisional and close encounter history, which could alter or destroy the binary orbit after its formation \citep{Nesvorny2011,Nesvorny2018,Parker2010,Parker2012a,Brunini2016}.
\\

\begin{figure}
        \centering
        \includegraphics[]{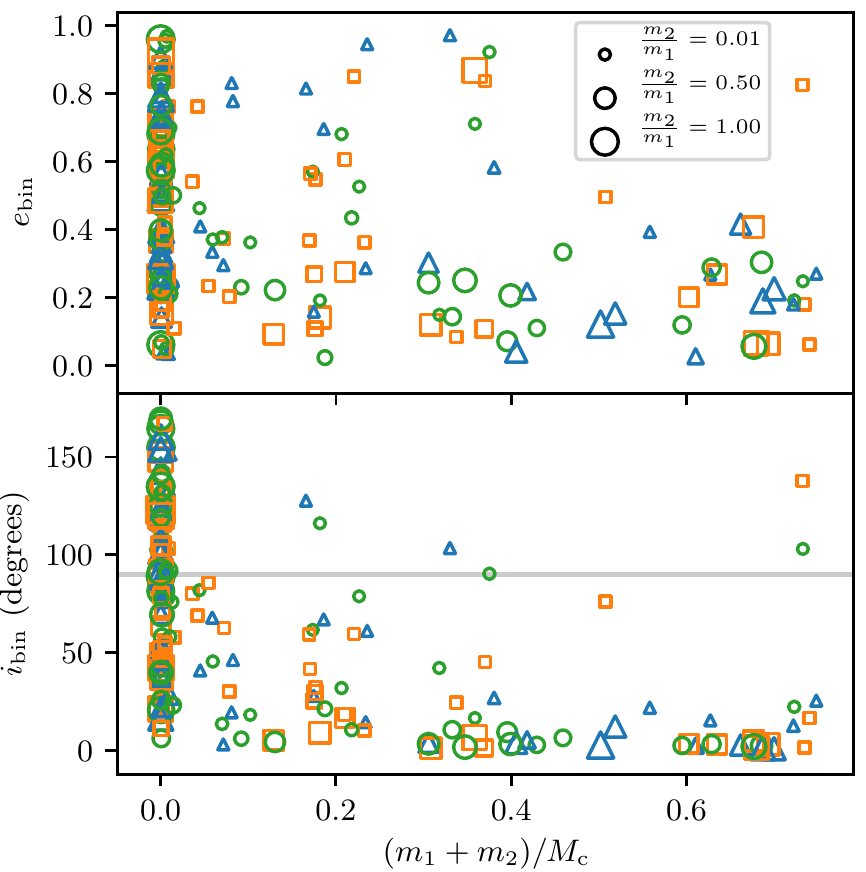}
        \caption{Relation between the binary orbital parameters (eccentricity and inclination) and the binary mass parameters. 
                The normalised system mass is shown on the $x$ -axis, and the mass ratio is represented by marker size.
                Marker shape and colour denote cloud mass (same as figure \ref{fig:binary_log_mass_ratios_mult}).}
        \label{fig:binary_orbits2}
\end{figure}

Figure \ref{fig:binary_orbits2} demonstrates how the binary orbital elements scale with the binary mass parameters.
There are a large number of high-eccentricity ($e_\mathrm{bin}\gtrsim0.5$) and retrograde ($i_\mathrm{bin}>90\degr$) orbits in our dataset, whereas in figure 5 of \citet[]{Nesvorny2010} there is a preference for lower eccentricities and only prograde systems.
In general, these more excited orbits occur for the lower mass binaries $(m_1+m_2)/M_\mathrm{c}\lesssim0.1$, that is,\ these are the additional atomistic binaries detected by the deeper orbit search in this work.
Furthermore, at higher system masses the high $e_\mathrm{bin}$ and $i_\mathrm{bin}$ systems have low mass ratios $m_2/m_1\lesssim0.1$, that is,\ they are the satellite systems where a small companion has been captured around a larger body.

For the complete dataset of 188 simulated binary orbits, $71\%$ of systems were prograde (median inclination = $32\degr$) and $29\%$ were retrograde (median inclination = $123\degr$).
In contrast to the observed TNBs of \cite{Grundy2019}, we found no obvious trend for wide binaries to have low inclinations, but as mentioned above we have presented a distribution for the time of formation in the rotating reference frame of the pebble cloud undergoing gravitational collapse.
When we considered only the simulated binary systems with $m_2/m_1>0.1$ and $(m_2+m_1)/M_\mathrm{c}>0.1$ (i.e.\ the observable binaries), all 35 are prograde, and these systems had a median inclination of $3.5\degr$ (figure \ref{fig:binary_inclination_polar}).
Furthermore, figures \ref{fig:binary_orbits} and \ref{fig:binary_orbits2} show that the observable binaries generally have only moderate eccentricities.
These properties agree with \cite{Nesvorny2010,Nesvorny2019}, in which gravitational collapse produces prograde systems with low inclination relative to the rotation of the cloud and $e_\mathrm{bin}<0.6$.

\section{Discussion}

\subsection{Origin of contact binaries}
\label{sec:FormationOfMU69}

Our results have shown that gravitational collapse is an extremely efficient method of producing binary systems with a range of properties.
\cite{Stern2019} noted that gravitational collapse easily forms the nearly equal size ratios of an Arrokoth-like object, with low merger speeds \citep{Nesvorny2010,Fraser2017}, albeit for higher mass systems \citep[see also][]{Umurhan2019,McKinnon2019}. 
The collisional origin of Arrokoth was then investigated further by \cite{McKinnon2020}, who showed that a low-velocity impact can indeed preserve the bi-lobate structure.
Furthermore, \cite{Grishin2020} demonstrated that dynamical evolution of an initially wide binary can lead to an Arrokoth forming gentle merger.
As pointed out by \cite{Nesvorny2018a} for the case of the comet 67P, \cite{Simon2017a} provided evidence that streaming instability may scale down to produce less massive $<100\, \si{km}$ objects.
High-resolution simulations of the streaming instability by \cite{Li2019} suggest that there could be a turnover in the mass distribution of clouds formed by streaming instability at $\sim5\e{17}\,\si{kg}$ (Appendix \ref{app:streaming_instability}), implying that clouds with masses lower than this may be less common.

In figure \ref{fig:binary_magnitudes} we demonstrated that gravitational collapse can produce both large and small binary systems from a single pebble cloud, including those that are equivalent to the masses of the components of Arrokoth.
Therefore we do not need to rely entirely on clouds formed by streaming instability scaling all the way down to masses similar to Arrokoth; with gravitational collapse small binaries can be produced in parallel to the larger TNBs.
It has also been pointed out that Arrokoth has an obliquity close to $90\degr$, and that it would be uncommon to produce such an inclined orbit through gravitational collapse \citep{McKinnon2019}.
As shown in figure \ref{fig:binary_orbits2}, the low-mass binary orbits in our dataset are not constrained to have low prograde binary orbits. 
Rather, our simulations produce just as many perpendicular orbit binaries as those with low inclinations for systems in the mass range of Arrokoth.

Regardless of the way it formed, a low-mass proto-Arrokoth binary system would have to subsequently evolve into a contact binary.
As discussed by \cite{Stern2019}, \cite{McKinnon2020} and \cite{Grishin2020}, the binary system would have had to lose angular momentum in order for the components to come together.
Proposed mechanisms include gas drag, the Kozai-Lidov mechanism, tidal evolution, and collisions.
Furthermore, this could possibly be achieved by ejection of another bound body, which is certainly possible given that gravitational collapse commonly produces high-multiplicity systems (section \ref{sec:further_dynamical_evolution}).

\subsubsection{Formation of contact binaries through gentle collisions}
\label{sec:contact_binary_collisions}

\begin{figure}
        \centering
        \includegraphics[]{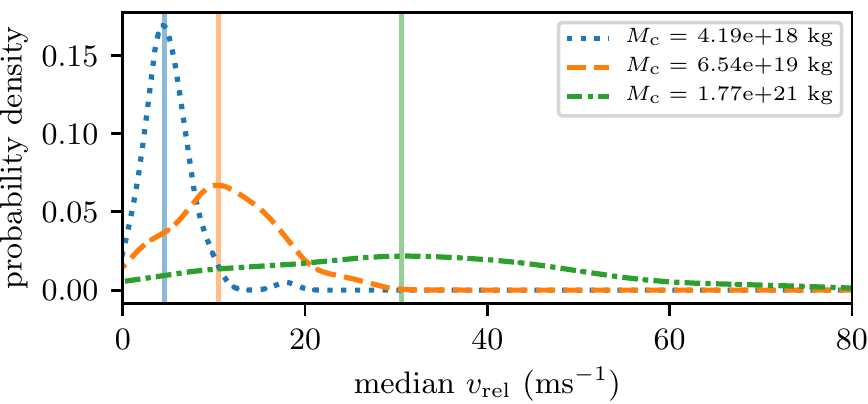}
        \caption{Probability density distribution of the median collisional velocity for all recorded collisions with $t>50\, \si{yr}$ for each gravitational-collapse simulation. Line style and colour denote cloud mass. The median value of each distribution is shown by a vertical line.
        }
        \label{fig:collision_median_velocity}
\end{figure}

We also considered the possibility that a contact binary forms directly through the collision of two bodies during cloud collapse \citep{McKinnon2019,McKinnon2020}.
In gravitational collapse the mean particle velocity in the collapsing cloud depends on cloud mass. 
In figure \ref{fig:collision_median_velocity} we present the distribution of median collisional velocities for all collisions recorded in the second half of all simulations ($t>50\, \si{yr}$) to gauge what the collisional environment in the cloud is like after the initial cloud collapse and particle growth phase.
The median collisional velocities are $\sim5, \, 10,\text{and} \, 30\, \si{m\, s^{-1}}$ for the low, intermediate, and high cloud mass, respectively.

If the collisional velocity is low enough then a merging collision may occur, possibly producing a contact binary similar to Arrokoth.
As mentioned in section \ref{sec:collision_resolution}, merging collisions are possible for $v_\mathrm{rel}<v_\mathrm{esc}$.
The density and mass of Arrokoth is unknown.
Using the observed volume, when we assume a characteristic cometary density of $500\, \si{kg\, m^{-3}}$ up to our fiducial $1000\, \si{kg\, m^{-3}}$, then the system mass is $m_\mathrm{A}=1$ -- $2\e{15}\, \si{kg}$.
The escape velocity is therefore $v_\mathrm{esc}=(2Gm_\mathrm{A}/r_\mathrm{A})^{1/2} = 4$ -- $6\, \si{m\, s^{-1}}$.
For Arrokoth, $v_\mathrm{esc}$ is comparable to the median collisional velocity of the lowest cloud mass, although $\sim17\%$ of collisions in the intermediate cloud mass also have $v_\mathrm{rel} \leq5\, \si{m\,s^{-1}}$.
When we consider only the collisions between particles with masses $m_1+m_2<m_\mathrm{A}$ , the median collisional velocities are even lower: $\sim 2$ and $6\, \si{m\, s^{-1}}$ for the low and intermediate cloud masses, respectively.
Therefore collisions in these clouds would frequently be merging for Arrokoth-mass particles.
The low cloud mass simulations are able to produce binary systems comparable to the mass of Arrokoth (figure \ref{fig:binary_magnitudes}), and therefore $M_\mathrm{c}\lesssim4.2\e{18}\, \si{kg}$ pebble clouds appear to be good candidates for producing an Arrokoth-like system through a gentle collisional merger.
We note, however, that for the intermediate and high cloud masses, particles as small as Arrokoth are beyond the numerical resolution of our simulations (see section \ref{sec:numerical_resolution}).

The detailed analysis of the collision of the Arrokoth components, for example,\ deformation of the bodies during collision, is beyond the scope of this paper.
However, smoothed particle hydrodynamics studies such as \cite{Jutzi2015} and \cite{Sugiura2018} indicate that there are impact parameters at these low velocities where collisions can retain the shapes of the colliding particles. 
Most recently, \cite{McKinnon2020} found that Arrokoth could have formed collisionally provided the impact was no greater than the escape velocity, approximately a few meters per second.
Furthermore, if streaming instability and gravitational collapse does indeed scale down to lower cloud masses to form Arrokoth, as implied by \citet{Stern2019}, from the trend in figure \ref{fig:collision_median_velocity} we would expect collisional velocity to also decrease.
The collisional environment in these low-mass clouds would then be even more conducive to the formation of a contact binary through a gentle merger.

\subsubsection{Ratio of binary to single planetesimals at low masses}
\label{sec:low_mass_single_to_bin}

In section \ref{sec:observable_binary_single_ratio} we discussed the number of bound systems brighter than magnitude 25 compared to singles of the same size that were produced by gravitational collapse.
We now consider the number of binaries in the mass range of Arrokoth produced by our simulations, with the assumption that such a binary may evolve into a contact configuration.
Figures \ref{fig:binary_log_mass_ratios_mult} and \ref{fig:particle_mass_dist} show that the distribution of binary system mass with mass ratio and the mass distribution of particles formed by gravitational collapse are similar for each initial cloud mass.
By normalising the particle masses with respect to the cloud mass, we see that the overall shape and structure of these distributions is invariant for the cloud masses investigated here.
We therefore assumed that these distributions would all scale down to a lower cloud mass of $M_\mathrm{c,low}=5\e{17}\, \si{kg}$, the most common cloud mass produced by the streaming instability simulations of \cite{Li2019}.
We used our 144 cloud dataset, normalised by the initial cloud mass, to predict the outcome of gravitational collapse for a cloud of mass $M_\mathrm{c,low}$.
Using the estimated mass for Arrokoth (section \ref{sec:contact_binary_collisions}), we considered all binaries within a fractional mass range of $2\e{-3}$ -- $4\e{-3} M_\mathrm{c}$.
We found that on average each cloud would produce $\sim0.07$ such binaries, compared to $0.64$ single particles of the same mass.
This means that for every Arrokoth-like binary created through the gravitational collapse of a $M_\mathrm{c,low}=5\e{17}\, \si{kg}$ mass cloud, we would expect to form $\text{about nine}$ single planetesimals in the same mass range.

Because the first cold classical object visited by a spacecraft is a contact binary, this may imply that a large fraction of small cold classicals could be contact binaries.
Although we produced a reasonable number of Arrokoth-mass binaries compared to singles, we cannot claim based on our estimate that all cold classical contact binaries started their lives as binaries formed by gravitational collapse.
Our predicted binary-to-single ratio is consistent with the estimate of \cite{Thirouin2019} that $\sim 10$ -- $20\%$ of the cold classical population could be contact binaries.
However, this is dependent on all binaries becoming contact systems, and as mentioned previously, we did not account for the evolution of the binaries after their formation.
In addition, our prediction for the binary-to-single ratio is highly dependent on the choice of $M_\mathrm{c,low}$.
We obtained this value from \cite{Li2019}, where the conversion from simulation mass units to physical mass units depended on the choice of protoplanetary disc model (Appendix \ref{app:streaming_instability}), let alone any uncertainties in the reported value.
For example, when we repeat the above analysis for $M_\mathrm{c,low} = 1\e{17}\, \si{kg}$, the binary-to-single ratio is 3:1 for Arrokoth masses.
This behaviour is a result of the steep power-law mass distribution and the break to a flat distribution as shown in figure \ref{fig:particle_mass_dist}. 
For a given cloud mass, when $m_\mathrm{A}$ lies below this break, a surplus of singles will be produced relative to the binaries.
An interesting implication of this is that if there is indeed a minimum cloud mass that can form through a streaming instability, and the particle mass distribution is invariant with cloud mass, then there is a critical mass below which nearly all planetesimals form as singles.

\subsection{Simulation limitations}
\label{sec:sim_limitations}

There are certain limitations to modelling gravitational collapse, as originally discussed by \citet{Nesvorny2010}.
The use of the particle inflation factor $f^*$ is necessary to increase the collision rate in the gravitational-collapse simulation.
In reality, the collision rate of the pebble cloud would be high due to the high number density of particles, but when we simulate gravitational collapse we are limited to using a smaller number of computational particles.
The use of $f^*$ can be problematic.
Increasing $f^*$ directly increases the collisional cross-sectional area of the particles and therefore affects the rate of collisions and mass accretion.
This is shown in figure \ref{fig:binary_mass_ratios_by_f}, where we found that the types (and frequency) of binaries formed through gravitational collapse were dependent on $f^*$.
In particular, we found that $f^*=30$ was the most efficient producer of binaries with a high system mass, high mass ratio.
We  followed \cite{Nesvorny2010} in only considering `intermediate' values of $f^*$, which strike the balance between no inflation and full inflation that conserves the total surface area of a realistic pebble cloud. Even within this range, $f^*$ can be tuned to produce binaries that do (or do not) match observations.
Additionally, as shown in figure \ref{fig:binary_rel_sep}, the use of size inflation severely limits the minimum separation of binary systems that can be formed.
Tight systems have a tendency to merge into a single planetesimal due to the size of the particles.
Furthermore, size inflation means that the particles are unrealistically large and the density is low.
Inflated particle density is given by $\rho^\prime=\rho/f^{*3}$, therefore for $f^*=100$, $\rho^\prime=\rho\e{-6}$!
Because of this effect, we were required to shrink particles to their physical radii in order to make a meaningful comparison to the observed systems (see figures \ref{fig:binary_size_ratios} and \ref{fig:binary_magnitudes}).

As mentioned in section \ref{sec:collision_resolution}, the use of $f^*$ can also affect collisional velocities in the cloud.
We detected a collision when \textit{\textup{artificially large}} inflated particles were overlapping, whereas in reality the particles would have had longer to accelerate towards each other before colliding at their true physical radii.
On the other hand, our numerical framework with a finite number of computational particles is biased to have higher collisional velocities than reality in any case. For systems with a fixed total mass, using a smaller number of more massive particles will generally result in stronger gravitational scattering events and therefore higher velocity excitation.       
All in all, these issues highlight the inherent problems when attempting to simulate a large number of collisionally interacting real particles, and should be borne in mind when interpreting these results.
\\

It has also been assumed that the effects of gas drag are negligible for these binary formation simulations.
The presence of gas is required for the initial formation of the cloud through streaming instability, but \cite{Nesvorny2010} showed that the collision timescale of a particle during gravitational collapse dominates the aerodynamic stopping time, therefore the effects of gas can be safely ignored.
In addition, they claimed that gas drag should not significantly affect the binary orbits over the timescale of the gas disc.
With the current interest in Arrokoth, \citet{McKinnon2019} and \citet{Umurhan2019} stated that this assumption may need to be revisited, especially when the evolution of binary systems into contact binaries is investigated.

\subsubsection{Cloud initial conditions}
\label{sec:cloud_initial_conditions}

The gravitational-collapse simulations may also be limited by the simplicity of the pebble cloud model.
We have followed \citet{Nesvorny2010} in using a uniform spherical approximation of the pebble cloud formed by streaming instability.
These conditions were originally chosen because the resolution of hydrodynamical simulations of the streaming instability is generally not high enough to resolve the properties of the extremely dense particle clumps.
In reality, the streaming instability is a more elaborate process, and the mass, velocity, and angular momentum distributions are more complex.
For example, we would expect that mass is generally more concentrated in the centre of the pebble cloud.
The streaming instability simulations of \cite{Nesvorny2019} and \cite{Li2019} produced particle clumps with a decreasing radial mass distribution within the Hill sphere of the clump.
Furthermore, the shape of the pebble cloud is likely stretched by the local shear forces and would be more ellipsoidal.
This increased central density could lead to faster collapse times in the centre of the cloud compared to the outer regions, and an ellipsoidal cloud would have more initial angular momentum, both of which could affect the properties of any binaries formed.
Future work that simulates gravitational collapse should consider investigating these more complex initial conditions.

High-resolution hydrodynamic simulations have shown that there are mass limits to the pebble clouds formed by streaming instability.
\cite{Nesvorny2019} reported an upper limit to the clouds that formed, and \cite{Li2019} provided evidence of a turnover at low clump masses (Appendix \ref{app:streaming_instability}).
For the high-resolution `Run I' in \cite{Li2019}, there is evidence of a turnover in the mass distribution of clumps produced by streaming instability at $5\e{17}\, \si{kg}$, where the distribution peaks.
The simulations of \cite{Li2019} and \cite{Nesvorny2019} suggested that the highest mass of a cloud that is produced by streaming instability is $2\e{20}$--$1\e{21}\, \si{kg}$.
In our simulations, the low and intermediate cloud masses lie between these lower and upper bounds, and we see now that the highest cloud mass appears to be unlikely to form by streaming instability.
Our results support an upper limit on cloud mass; we have shown in section \ref{sec:binary_magnitudes} that the high cloud mass produces an over-abundance of observable single planetesimals and also binary systems that are not observed.
Given the turnover and peak in frequency around $M_\mathrm{c}=5\e{17}\, \si{kg,}$ it would be wise in future work on gravitational collapse to probe lower cloud masses.

\subsubsection{Numerical resolution}
\label{sec:numerical_resolution}

The number of simulation particles, $N_\mathrm{p}=10^{5}$, limited the mass of the smallest particles we could investigate for a particular initial cloud mass.
Because of the computational limits for an $N$-body simulation with interactions between all particles and accurate collision detection, $N_\mathrm{p}=10^{5}$ was the highest resolution that was feasible for this work at the time. 
This is particularly important when we consider objects on Arrokoth-mass scales.
For the low, intermediate, and high cloud mass, a particle of mass $m_\mathrm{A}=2\e{15}\, \si{kg}$ would be composed of $n=47,3,\text{and }0.1$ simulation particles (each of mass $m_0=M_\mathrm{c}/N_\mathrm{p}$), respectively.
It is clearly impossible to form such a system with a high cloud mass simulation, and for the intermediate cloud mass $m_\mathrm{A}$ is below the orbit search limit of $4m_0$ (section \ref{sec:OrbitSearchAlgorithm}).
Furthermore, we are most interested in objects that have had a reasonable number of collisions, and have undergone collisional evolution due to the process of gravitational collapse, as these are the objects that are most likely to host a bound system.
Therefore, only the low cloud mass simulations are suitable for probing such low masses.

As discussed in sections \ref{sec:contact_binary_collisions} and \ref{sec:cloud_initial_conditions}, the lower collisional velocities and the distribution of cloud masses formed by streaming instability are good reasons to investigate the gravitational collapse of lower cloud masses.
We have shown that the process of binary formation through gravitational collapse is generally invariant with initial cloud mass, for example\
the mass distribution of particles (figure \ref{fig:particle_mass_dist}), the binary system mass and mass ratio distribution (figure \ref{fig:binary_log_mass_ratios_mult}), and the binary orbital properties (figures \ref{fig:binary_inclination_polar} and \ref{fig:binary_orbits2}), therefore we would expect the results for lower cloud masses to be largely the same as well.
We would expect such simulations to form a higher number of binaries of mass comparable to Arrokoth (section \ref{sec:low_mass_single_to_bin}), and moreover, $m_0$ would be a much smaller fraction of $m_\mathrm{A}$. 
Therefore such binaries would be better resolved as they must undergo more collisions to form.

\subsubsection{Completeness of binary identification}
\label{sec:orbit_detection}

Our results may be limited by our methods of searching for bound particles.
Only the $N_\mathrm{lim}=100$ most massive particles were searched for orbits, as described in section \ref{sec:OrbitSearchAlgorithm}, therefore we may have missed a small fraction of bound systems.
This cut was made because after gravitational collapse  a single particle typically dominated in mass (figure \ref{fig:particle_mass_dist}), which may be a host to bound companions.
Our orbit search was designed primarily to find this case, but also any additional bound systems of significant mass (the more massive particles are most likely to host companions).
We may miss some systems, especially if they are of low mass, which is particularly relevant for candidate Arrokoth-like systems.
When we tested searching a larger number of particles ($N_\mathrm{lim}=200$) only a small number of additional orbits were found, and we always detected the high system mass and mass ratio systems, as expected.
Another issue is that thus far we only considered the orbits found at the end of the simulation, at $t=100\, \si{yr}$. 
However, there could be a number of lower mass binaries that are ejected from the simulation box at earlier times.

Furthermore, we chose to search for bound orbits between pairs of particles, as opposed to assessing the total energy of clumps of particles for binding. This means that we could have missed particles that are not bound to the primary particle, but are bound to the centre of mass of the clump. Any particles that we miss would most likely be only weakly bound as they would either be distant or have a high velocity relative to the primary. We chose our method for its simplicity; structure-finding in $N$-body simulations is a complex problem, and the results may depend on the choice of algorithm \citep{Knebe2013}. The pair-wise orbit search was suitable in our analysis as this method generally identifies high-multiplicity systems where one or more components are only instantaneously bound. 
During the $10^4$ years of further dynamical evolution the multiplicity of these systems was always drastically reduced, with the majority of systems evolving to simple binary pairs. Therefore we would not expect additional loosely bound particles (that may be detected when the system energy is considered) to survive this further evolution either.

On the other hand, we investigated the types of binaries that are produced by gravitational collapse in more detail than previous work by performing a deeper orbit search.
We included all bound systems that were detected by the orbit-search algorithm, whereas \citet{Nesvorny2010} included only the single most massive system produced by each cloud.
Furthermore, figure \ref{fig:binary_log_mass_ratios_mult} showed that in general the detected systems were well above the detection limits of the orbit search.
Using this deep orbit search, we found that it is common for each collapsing cloud to produce several bound systems.
We ran 144 gravitational-collapse simulations and detected a total of 287 bound systems of particles at $t=100\, \si{yrs}$.
After $10^{4}\, \si{yr}$ of dynamical evolution we were left with 223 systems; each cloud produced $\sim1.5$ bound systems on average.
Gravitational collapse is therefore an extremely efficient mechanism for producing bound systems of planetesimals.

\subsubsection{Collision detection and mass accretion}
\label{sec:timestep_collisions}

\begin{figure}
        \centering      
        \includegraphics[]{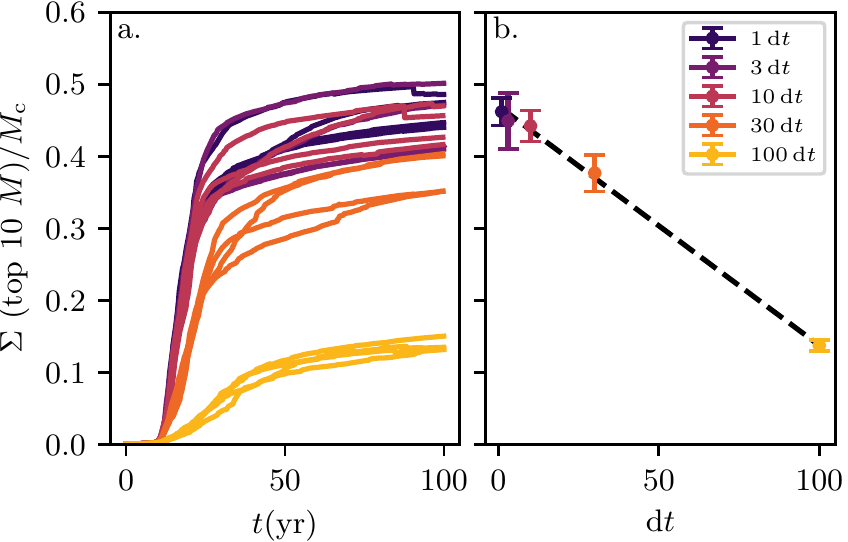}
        \caption[]{Effect of changing the timestep on mass accretion in a test set of simulations with parameters $f^*=30$, $M_\mathrm{c}=6.54\e{19}\, \si{kg}$, $X=0.75,$ and four unique seed positions.
                This set of four simulations was repeated with the timestep $\mathrm{d}t$ increased by factors 1, 3, 10, 30, and 100, as indicated by line and marker colour in both panels.
                Panel a shows the total mass of the ten most massive particles in each simulation as a function of time. 
                Panel b shows the total mass accreted into the ten most massive particles at $t=100\, \si{yr}$, where the error bar markers indicate the mean and standard deviation of each set of four simulations.
                As the timestep is increased, the mass accretion drops in an approximately linear fashion (dashed line).
        }
        \label{fig:mass_dt}
\end{figure}

In section \ref{sec:binary_sizes} we found that our simulations generally produced more massive binaries than those of \cite{Nesvorny2010}. 
This would imply that particles in our simulations underwent more merging collisions than in previous work.
In order to test this, we investigated how the simulation timestep affects the mass accretion of the most massive particles in the cloud.
A sample of four cloud-collapse simulations (parameters: $f^*=30$, $M_\mathrm{c}=6.54\e{19}\, \si{kg}$, $X=0.75,$ and four seed positions) were repeated with the timestep $\mathrm{d}t$ increased by factors of 1, 3, 10, 30, and 100.
We tracked the total mass of the ten most massive particles as a function of simulation time, as shown in figure \ref{fig:mass_dt}.a.
Considering the rate of mass accretion and comparing this to similar growth tracks of the \cite{Nesvorny2010} simulations \citep[presented in supplementary figure 4 of][]{Fraser2017}, we see that in both cases the highest mass is accumulated early in the simulation, after which mass accretion levels off.
This initial period of rapid runaway growth typically occurs until $t\sim 2 t_\mathrm{ff}$. 
During this time, the rate of mass accretion is dependent on particle mass $m$ such that $\mathrm{d}m/\mathrm{d}t \propto m^a$ and $a>1$. After this time, the cloud has begun to reach a post-collapse equilibrium state where collisions are less frequent (e.g.\ figure \ref{fig:collision_criterion}); the rate of mass accretion slows down and particle growth is no longer runaway.

In figure \ref{fig:mass_dt}.b we consider the total mass of the ten most massive particles at $t=100\, \si{yr}$. It is clear that as the timestep is increased potential collisions are missed, and mass accretion amongst the largest particles is less efficient.
The collision criterion (section \ref{sec:timestep}) is no longer satisfied.
It is therefore possible that the gravitational-collapse simulations of \cite{Nesvorny2010} may have missed some collisions as a result of using a large timestep and straight line extrapolation collision detection \citep{Richardson2000}.
As described by \cite{Rein2011}, particle trajectories in Hill's approximation are curved and not straight.
 \cite{Rein2011} therefore stated that `it is better to detect collisions approximately along exact trajectories than the reverse'.
In our simulations, we have been able to do this by conducting more computationally intensive overlapping particle collision detection, and we tested the collision resolution criterion for all simulations, as described in section \ref{sec:timestep}.

Interestingly, it would appear that the differing amounts of mass accretion between the two works have not significantly affected the main results.
Both our work and that of \cite{Nesvorny2010} successfully produce high-mass ratio binaries that have similar properties to the observed TNBs: low to moderate eccentricity, low prograde inclinations, and separations ranging $\sim 10^3 - 10^8\,\si{km}$.
The only difference is that our binary system masses are systematically more massive than those of \cite{Nesvorny2010}, which offers another explanation for why our results favour the collapse of low-mass clouds (figure \ref{fig:binary_magnitudes}).

\section{Conclusions}

We have investigated in detail the formation of transneptunian binaries through the gravitational collapse of a pebble cloud.
This is the first independent investigation into this mechanism since it was originally presented by \cite{Nesvorny2010}.
Our main results are summarised below.

\begin{enumerate}
        \item Gravitational collapse is extremely efficient at producing bound systems, frequently producing multiple bound systems per cloud.
        \item We reproduce the main findings of \cite{Nesvorny2010}; gravitational collapse produces binaries with similar mass, wide separation, moderate eccentricity, and low prograde inclination.
        \item By performing a deeper orbit search, we found that a range of additional binaries were produced by each cloud; these span a range of system masses, mass ratios, and binary orbital properties.
        \item Our results support the upper limit to cloud mass reported in high-resolution streaming instability simulations; the highest cloud mass in our simulations produces binaries that have no observational counterpart, and it is also inconsistent with all planetesimals forming as binaries.
        \item Gravitational collapse can produce systems that might evolve into Arrokoth-like objects, with low system mass, low-velocity merging collisions, and high-obliquity binary orbits.
\end{enumerate}

Our study reproduces and supports some of the results of \cite{Nesvorny2010}; gravitational collapse can form binary systems that have nearly equal-size components, prograde low-inclination, and moderate-eccentricity orbits.
When compared to the latest TNB data, we find that we can produce binary systems with similar properties to what is observed.
In our dataset, observations are best matched by gravitational collapse of a low-mass cloud.
Furthermore, by performing a deeper orbit search on particles in the collapsed cloud, we have found that a single cloud can produce more than one bound system.
These systems display a diverse range of properties when system masses, mass ratios, binary orbital elements, and multiplicity are considered.
We have shown that gravitational collapse is an extremely efficient mechanism for producing bound planetesimal systems.

The collapse of high-mass clouds ($M_\mathrm{c}=1.77\e{21}\, \si{kg}$) produces large equal-mass binaries for which there is no observational equivalent. This disfavours formation of TNBs through collapse of high-mass clouds and supports the simulations of \cite{Nesvorny2019}, which constrain the upper mass of clouds formed by streaming instability.
This is reinforced by high-mass clouds producing an excess of single planetesimals.
Only the collapse of a low-mass cloud can produce all planetesimals as binaries, a requirement of \cite{Fraser2017}.
Moreover, when we consider the distribution of binary orbital elements in our dataset (including those that are currently outside the detectable limits), we produce a much wider range of eccentricity and inclination than what is seen in the TNBs observed today. 
These results could be used in future work to constrain the evolution of binary systems from formation until the present day, testing models of KCTF, collisions, encounters, and migration.

Importantly, by performing a deeper orbit search we find binary systems down to very low masses, typical of objects such as 2014 MU$_{69}$ Arrokoth.
We show that a low-mass collapsing cloud has a favourable low-velocity environment for merging collisions that may preserve the shape of components.
This lends support to theories of formation of contact binaries from a binary system formed by gravitational collapse, but this requires further investigation into the evolution of TNBs over longer timescales and detailed study into collisions between components, which was recently addressed by \cite{McKinnon2020}.
In addition, the low-mass binaries in our dataset are not constrained to the same inclination distribution as the systems found by \cite{Nesvorny2010}, which perhaps offers a route of forming an extremely oblique Arrokoth.

\begin{acknowledgements}
We thank the anonymous referee for their considered review and useful comments which improved this work.
  J.R. acknowledges funding from the Northern Ireland Department for Education.
  W.F., A.F. and P.L. acknowledge support from Science and Technology Facilities Council grant ST/P0003094/1.
  J.R. also thanks the LSSTC Data Science Fellowship Program, which is funded by LSSTC, NSF Cybertraining Grant \#1829740, the Brinson Foundation, and the Moore Foundation; his participation in the program has benefited this work.
  
  Simulations in this paper made use of the \rebound code which is freely available at \url{http://github.com/hannorein/rebound}.
  J.R. is grateful for the assistance of Hanno Rein in developing the code used in this study, through his active support of \rebound users.
  J.R. thanks David Nesvorn\'{y} in particular, for providing valuable insight and guidance throughout the project, as well as Will Grundy for helping to interpret TNB observations.
  J.R. is also appreciative the support of the Astrophysics Research Centre QUB, in particular Tom Seccull and Richard Smith.
  
  This research made extensive use of the HPC facilities of Queen's University Belfast, and also the facilities of the Canadian Astronomy Data Centre operated by the National Research Council
  of Canada with the support of the Canadian Space Agency. 

  The following software packages were also used in this work: \texttt{matplotlib} \citep{Hunter2007}, \texttt{numpy} \citep{Oliphant2006,VanDerWalt2011}, \texttt{scipy} \citep{Virtanen2020}, \texttt{scikit-learn} \citep{Pedregosa2011} and \texttt{pandas} \citep{McKinney2010}.
  
\end{acknowledgements}

%
\bibliographystyle{aa} 
\bibliography{grav_collapse_paper_library.bib} 


\begin{appendix}

\section{Calculating magnitudes}
\label{app:magnitudes}

We calculate an apparent magnitude (in waveband $\lambda$) of an object of radius $r$ $\si{km}$ and geometric albedo $p_\lambda$ using the equations in \cite{Petit2008} and \cite{Noll2008},
\begin{align}
\mathrm{m}_\lambda&=\mathrm{m}_{\sun,\lambda} -2.5\log\left(\frac{p_\lambda r^2 \phi(\alpha)}{2.25\e{16} R^2 \Delta^2}\right)\\
&=5\log\left(\sqrt{2.25\e{16}} 10^{0.2\mathrm{m}_{\sun,\lambda}}\frac{ R \Delta }{\sqrt{p_\lambda} r }\right)\\
&=5\log\left(C_\lambda \frac{R \Delta}{\sqrt{p_\lambda} r}\right),
\end{align}

where for the $V$ band, $\mathrm{m}_{\sun,V}=-26.76$ \citep{Willmer2018} is the $V$ -band solar magnitude, $C_V=664.5\, \si{km}$ \citep{Noll2008}, and we have assumed that the phase function $\phi=1$ (i.e.\ phase angle $\alpha=0$) and the object has no rotational variation.
$R$ is the heliocentric distance of the object and $\Delta=R- 1$ is the geocentric distance for $\alpha=0$, both in $\si{AU}$.

The magnitude difference between a binary with components of size  $r_1$ and $r_2$ is then
\begin{equation}
\Delta \mathrm{m}_\lambda = \mathrm{m}_{\lambda,2} - \mathrm{m}_{\lambda,1} =  5\log\left(\frac{r_1}{r_2}\sqrt{\frac{p_{\lambda,1}}{p_{\lambda,2}}}\right) \label{eqn:delta_mag}
.\end{equation}

\section{Streaming instability masses}
\label{app:streaming_instability}

\cite{Nesvorny2019} reported a mass distribution of clouds formed by streaming instability.
In their supplementary information, they discussed the highest planetesimal mass that can be produced by streaming instability and the conversion between simulation mass units and physical mass units, which requires a disc temperature.
The maximum cloud mass produced by their runs A and C is
\begin{equation}
M_\mathrm{max} \sim0.25 - 0.75 T_{25}^{3/2} M_\mathrm{Ceres}
,\end{equation}
where $M_\mathrm{Ceres}=9.4\e{20}\, \si{kg}$ is the mass of Ceres and $T_{25}=T/(25\, \si{K})$ for disc temperature $T$.
A fully flared disc model predicts a temperature profile of exponent $-3/7$ \citep{Chiang2010}. 
Therefore \citet{Nesvorny2019} used
\begin{equation}
T=39R_{45}^{-3/7}
,\end{equation}
where $R_{45}=r/(45\,\si{AU})$ and $r$ is the radial heliocentric distance. 

Using these relations, for $r=45\, \si{AU}$, $T=39\, \si{K}$, $M_\mathrm{max}\sim4.6\e{20}$--$1.4\e{21}\, \si{kg}$.
For $r=30\, \si{AU}$, $T=46\, \si{K}$, $M_\mathrm{max}\sim5.9\e{20}$--$1.8\e{21}\, \si{kg}$.
Therefore we assume a maximum cloud mass of $\sim1\e{21}\, \si{kg}$.
\\

In their high-resolution streaming instability simulations, \cite{Li2019} found evidence for a turnover at low mass in the distribution of cloud masses produced by streaming instability.
They analysed two simulations, Run I and Run II (equivalent to the \cite{Nesvorny2019} runs A and C respectively).
Both have the following dimensionless simulation parameters: global radial pressure gradient parameter $\Pi =0.05$, and relative strength of particle self-gravity to the tidal shear $\tilde{G}=0.05$. 
Run I has a stopping time and gas-to-surface density $(\tau_s,Z)=(2.0,0.1)$ and is higher resolution $(\Delta x = H/5120)$ in a smaller domain.
Run II has $(\tau_s,Z)=(0.3,0.02)$ and is lower resolution (half of Run I) over a larger domain.
The dimensionless simulation mass units for the runs correspond to physical masses $M_G=0.19$ and $0.0015\, M_\mathrm{Ceres}$ , respectively.
As for \cite{Nesvorny2019} above, these values are dependent on disc conditions, such as disc radius $R$ and temperature profile $T\propto R^{-3/7}$.

The high resolution of Run I allows the production of lower cloud masses than Run II, thus allowing investigation of the low end of the cloud mass distribution.
Figure 4 of \cite{Li2019} provides evidence of a turnover in the mass distribution below a peak at $\sim0.003\, M_G = 5.7\e{-4}\, M_\mathrm{Ceres}=5.4\e{17}\, \si{kg}$.
These results imply that clouds of mass $5.4\e{17}\, \si{kg}$ should be the most common clouds produced by streaming instability.

\section{Supplementary animation}

We include an animation showing the complete process of gravitational collapse\footnote{\texttt{doi.org/10.17034/6f4b3d90-c3ba-4510-add5-69e504480a74}} for a cloud with the parameters $f^*=30$, $X=0.5$ and $M_\mathrm{c}=6.54\e{19}\, \si{kg}$. 
The left-hand panel shows the $xy$ projection of the particle position in the simulation box.
The marker size scales linearly with particle radius $r_f$ , and the colour scales logarithmically with particle mass $m$ according to $m_\mathrm{rel}=\log (m / m_0)/\log (M_\mathrm{c} / m_0)$.
The panel is initially zoomed in on the cloud and then zooms out until it shows the extent of the whole simulation box.
The centre of mass of the cloud is marked with a black $\text{cross}$ and the position of the most massive particle is shown with a red $\text{plus}$.
The right-hand panel shows an expanded view centred on the most massive particle at that timestep. Here the marker size scales with $r_f^2$ to emphasise the larger particles.
The blue ellipse indicates the binary orbit that is detected between the most massive particle and its largest companion.

\end{appendix}

\end{document}